\documentclass[11pt]{article}

\usepackage[usenames,dvipsnames]{xcolor}
\definecolor{Gred}{RGB}{219, 50, 54}
\definecolor{ToCgreen}{RGB}{0, 128, 0}

\usepackage[margin=1in]{geometry}
\usepackage[T1]{fontenc}

\usepackage[scale=0.97]{XCharter} 

\usepackage[libertine,bigdelims,vvarbb,scaled=1.05]{newtxmath} 

\usepackage{babel}
\usepackage[spacing=true,kerning=true,babel=true,tracking=true]{microtype}

\DeclareMathAlphabet{\pazocal}{OMS}{zplm}{m}{n} 
\renewcommand{\mathcal}[1]{\pazocal{#1}}

\usepackage{makecell}
\usepackage{dsfont}

\usepackage{multirow}
\usepackage{amsmath,amsthm}
\usepackage{bm}
\usepackage{bbm}
\usepackage{textgreek}
\usepackage{mathtools}
\usepackage{enumitem}
\usepackage[numbers,comma,sort&compress]{natbib}
\usepackage{authblk}
\usepackage{graphicx}
\usepackage[font=small]{caption}
\usepackage[labelformat=simple]{subcaption}

\usepackage{float}
\usepackage[linesnumbered,ruled,vlined]{algorithm2e}
\SetKwInput{KwInput}{Input}
\SetKwInput{KwOutput}{Output}
\SetKwInOut{Promise}{Promise}
\SetKwInput{Goal}{Goal}
\SetKwProg{Fn}{function}{}{}
\SetKwFor{RepTimes}{repeat}{times}{}
\SetKwFunction{Ver}{Verify}
\SetKwFunction{Prep}{PrepareState}
\usepackage{algorithmic}

\usepackage{physics}
\usepackage{footnote}
\usepackage{xcolor}
\usepackage{mathrsfs}
\usepackage{bbm}
\usepackage{braket}
\usepackage[colorlinks]{hyperref}
\usepackage{cleveref}
\hypersetup{
      colorlinks=true,
  citecolor=ToCgreen,
  linkcolor=Sepia,
  filecolor=Gred,
  urlcolor=Gred
  }
\usepackage{placeins}

\newtheorem{theorem}{Theorem}
 
\newtheorem{definition}{Definition}
\newtheorem{lemma}{Lemma}
\newtheorem{claim}{Claim}

\newtheorem{problem}{Problem}

\newcommand{\algo}[1]{\hyperref[algo:#1]{Algorithm~\ref*{algo:#1}}}

\newcommand{\E}{\mathbb{E}}

\newcommand{\cM}{\mathcal{M}}

\newcommand{\cD}{\mathcal{D}}

\newcommand{\cT}{\mathcal{T}}
\newcommand{\cA}{\mathcal{A}}

\usepackage{textgreek}

\DeclareMathOperator{\poly}{poly}

\DeclareMathOperator{\Var}{Var}

\DeclareMathOperator{\med}{median}

\renewcommand{\tr}{\mathrm{tr}}

\newcommand{\SWAP}{\textnormal{SWAP}}

\def\:{\hbox{\bf:}}

\renewcommand{\epsilon}{\varepsilon}
\allowdisplaybreaks

\begin{document}
%%%%%%%%%%%%%%%%%%%%%%%%%%%%%%%%%%%%%%%%%%%%%%%%%%%%%%%%%%%%%

%%%%%%%%%%%%%%%%%%%%%%%%%%%%%%%%%%%%%%%%%%%%%%%%%%%%%%%%%%%%%
% Make title
\title{On the sample complexity of purity and inner product estimation}

% \author{Anonymous Authors}
\author{Weiyuan Gong
\thanks{SEAS, Harvard University. Email: \href{mailto:wgong@g.harvard.edu}{wgong@g.harvard.edu}.}
\qquad\qquad
Jonas Haferkamp
\thanks{SEAS, Harvard University. Email: \href{mailto:jhaferkamp42@gmail.com}{jhaferkamp42@gmail.com}.}
\qquad\qquad
Qi Ye
\thanks{IIIS, Tsinghua University. Email: \href{mailto:yeq22@mails.tsinghua.edu.cn}{yeq22@mails.tsinghua.edu.cn}.}
\qquad\qquad
Zhihan Zhang
\thanks{IIIS, Tsinghua University. Email: \href{mailto:zhihan-z21@mails.tsinghua.edu.cn}{zhihan-z21@mails.tsinghua.edu.cn}.}
}
\date{\today}
\maketitle

\begin{abstract}
We study the sample complexity of the prototypical tasks \emph{quantum purity estimation} and \emph{quantum inner product estimation}. The former has been a key subroutine for experimental benchmarking and entropy estimation, while the latter is a core task for cross-platform verification that compares the unknown quantum states produced by different physical platforms. In purity estimation, we are to estimate $\tr(\rho^2)$ of an unknown quantum state $\rho$ to additive error $\epsilon$. Meanwhile, for quantum inner product estimation, Alice and Bob are to estimate $\tr(\rho\sigma)$ to additive error $\epsilon$ given copies of unknown quantum state $\rho$ and $\sigma$ using classical communication and restricted quantum communication. 

In this paper, we show a strong connection between the sample complexity of purity estimation with bounded quantum memory and inner product estimation with bounded quantum communication and unentangled measurements. Specifically, we prove the following results:
\begin{itemize}[leftmargin=*]
    \item \textbf{Sample complexity upper bound}:  We propose a protocol that solves quantum inner product estimation with $k$-qubit \emph{one-way} quantum communication and unentangled local measurements using $O(\med\{1/\epsilon^2,2^{n/2}/\epsilon,2^{n-k}/\epsilon^2\})$ copies of $\rho$ and $\sigma$. Our protocol can be modified to estimate the purity of an unknown quantum state $\rho$ using $k$-qubit quantum memory with the same sample complexity.
    \item \textbf{Sample complexity lower bound}: We prove that arbitrary protocols with $k$-qubit quantum memory that estimate purity to error $\epsilon$ require $\Omega(\med\{1/\epsilon^2,2^{n/2}/\sqrt{\epsilon},2^{n-k}/\epsilon^2\})$ copies of $\rho$. 
    This indicates the same sample complexity lower bound for quantum inner product estimation with \emph{one-way} $k$-qubit quantum communication and classical communication, and unentangled local measurements. 
    For purity estimation, we further improve the sample complexity lower bound to $\Omega(\max\{1/\epsilon^2,2^{n/2}/\epsilon\})$ for any protocols using an identical single-copy projection-valued measurement.
\end{itemize}
Additionally, we investigate a decisional variant of quantum distributed inner product estimation without quantum communication for mixed state and provide a lower bound on the sample complexity.
\end{abstract}

% \clearpage

% \tableofcontents
% \clearpage

%%%%%%%%%%%%%%%%s%%%%%%%%%%%%%%%%%%%%%%%%%%%%%%%%%%%%%%%%%%%%%

\newpage

\section{Introduction}
In recent years, notable progress has been made in the development of noisy intermediate-scale quantum devices utilizing different physical platforms~\cite{preskill2018quantum}. Benchmarking the physical systems on these quantum devices is of fundamental importance for verifying the computational outputs~\cite{eisert2020quantum}. However, the increasing size of quantum devices has pushed the standard verification methods such as state tomography~\cite{odonnell2016efficient,haah2016sample} and state certification~\cite{buadescu2019quantum} to the limit of their capabilities due to the exponential scaling of resources requirement in system size $n$. These certification algorithms focus on the task of comparing the noisy quantum state (process) with a known target. 
It is therefore desirable to develop protocols with feasible
resource requirements for cross-platform verification~\cite{elben2020cross}.
Such a protocol should directly compare unknown quantum states generated by different devices, which is an essential task in quantum benchmarking. 

In this work, we focus on the prototypical task of \emph{quantum inner product estimation}. Here, Alice and Bob are given copies of possibly different quantum states $\rho$ and $\sigma$, respectively. Their goal is to estimate the inner product $\tr(\rho\sigma)$ up to additive error $\epsilon$ with a high probability. 
%This task arose originally as a key subroutine of cross-platform verification~\cite{elben2020cross}, where Alice and Bob can be two quantum computers on different platforms. 
When Alice and Bob are able to share arbitrary quantum communication channels, a joint SWAP test suffices to efficiently estimate $\tr(\rho\sigma)$. However, quantum communication between Alice and Bob is limited in practice due to hardware constraints. In the extreme case when Alice and Bob are restricted to local quantum operations and classical communication (LOCC), this task is known as \emph{distributed quantum inner product estimation}~\cite{anshu2022distributed}. It was shown in the same work that $\Theta(2^{n/2})$ copies of $\rho$ and $\sigma$ are sufficient and necessary to solve distributed inner product estimation to constant accuracy $\epsilon$ even if joint local quantum measurements are allowed. 

Recent experimental advances in realizing quantum local area networks~\cite{kurpiers2018deterministic,magnard2020microwave,alshowkan2021reconfigurable} enable cross-platform verification schemes with quantum communication channels of small capacity~\cite{knorzer2023cross}. However, the capacities of these quantum wires between different platforms are restricted away from the $n$-qubit quantum communication channel that is required for joint SWAP tests. From the complexity perspective, it is desirable to characterize the sample complexity in terms of quantum communication~\cite{anshu2023survey}. 
Formally, we investigate the complexity of the following task:

\begin{problem}[Inner product estimation with \emph{$k$-qubit one-way quantum communication}]\label{prob:inner_product_k}
Alice and Bob are given $N$ copies of unknown quantum state $\rho$ and $\sigma$, respectively. In  each round of measurement, Alice and Bob each get access to one copy of $\rho$ and $\sigma$, and are able to perform $k$-qubit one-way quantum communication and any one-way classical communication. The goal is to estimate the quantum inner product $\tr(\rho\sigma)$ within additive error $\epsilon$.
\end{problem}

\noindent Here, we focus on the scheme where Alice and Bob are only allowed to perform \emph{unentangled (i.e., single-copy) measurements locally} as it is enough to provide the optimal sample complexity of $O(2^{n/2})$ when $k=0$ for distributed quantum inner product estimation~\cite{anshu2022distributed} and local joint measurements are relatively impractical in experiments. In addition, we assume that the communication between Alice and Bob is restricted to a one-way communication from Alice to Bob. The setting where Alice and Bob are allowed to have arbitrary $k$-qubit quantum communication and classical communication is beyond the scope of this work and remains an open question.

We demonstrate strong connections between the sample complexity of \emph{purity estimation} and quantum inner product estimation in different measurement schemes. Purity estimation, which estimates the purity $\tr(\rho^2)$ to additive error $\epsilon$ given copies of $\rho$, emerges as a key subroutine in many quantum information problems such as quantum benchmarking~\cite{eisert2020quantum} and experimental estimation of quantum R\'{e}nyi entropy on near-term devices~\cite{bluvstein2024logical}. As pointed out by~\cite{anshu2022distributed}, a single-copy protocol for distributed inner product estimation, which corresponding to \Cref{prob:inner_product_k} with $k=0$, indicates a single-copy protocol for purity estimation. Furthermore, by setting $\rho=\sigma$, we can observe that any algorithm for \Cref{prob:inner_product_k} indicates an algorithm for distributed estimation of $\tr(\rho^2)$ with bounded one-way quantum communication between Alice and Bob. It is thus natural to ask if we can show connections of sample complexity between \Cref{prob:inner_product_k} and purity estimation setting without communications.

We answer this question affirmatively by showing that the same sample complexity scaling for quantum inner product estimation with $k$-qubit one-way quantum communication (\Cref{prob:inner_product_k}) and purity estimation with \emph{$k$-qubit quantum memory}, which is defined as the following:
\begin{problem}[Purity estimation with $k$-qubit quantum memory, informal, see \Cref{def:learning_k}]\label{prob:purity_k}
Given $N$ copies of the unknown quantum state $\rho$, the goal is to estimate $\tr(\rho^2)$ within additive error $\epsilon$. The algorithm can take two copies of $\rho$ at a time, and maintain $n$ working qubits and a $k$-qubit quantum memory to store the data. In each round, the algorithm first makes a quantum measurement on the first copy and store $k$ qubits of the post-measurement state in the quantum memory. The algorithm then performs a quantum measurement on the $n+k$ qubits on the second copy and the quantum memory. 
\end{problem}
\noindent We note that, in contrast to the definition of learning with $k$-qubit quantum memory in Refs.~\cite{chen2022exponential,chen2024optimal} that can take $N$ oracle access to $\rho$ and store data in the quantum memory before the final measurement on $\rho^{\otimes N}$, the definition here follows the definition of learning with $2$-copy measurements and $k$-qubit quantum memory in Ref.~\cite{chen2024optimal}, which only allows the usage of at most \emph{two copies of $\rho$} in each round. This measurement setting is more feasible in experiment, and possesses the connection to the task of inner production estimation with unentangled local measurements between \emph{two} platforms.

In \Cref{sec:purity_to_ip}, we show that any protocol for inner product estimation where Alice and Bob share $k$ qubits of one-way quantum communication and any one-way classical communication (\Cref{prob:inner_product_k}) indicates a protocol for purity estimation with $k$-qubit quantum memory (\Cref{prob:purity_k}) with the same sample complexity scaling. Therefore, inversely, the lower bound for purity estimation with $k$-qubit quantum memory indicates a lower bound of the same scaling for inner product estimation where Alice and Bob share $k$ qubits of one-way quantum communication and any one-way classical communication. Using this connection, our first main result gives an algorithm to \Cref{prob:inner_product_k} and thus \Cref{prob:purity_k}:
\begin{theorem}[Informal, see \Cref{thm:partial_swap}]\label{thm:partial_swap_informal}
Given unknown quantum states $\rho$ and $\sigma$, and error parameter $\epsilon$, there exists a non-adaptive algorithm that solves inner product estimation of $\tr(\rho\sigma)$ with $k$-qubit one-way quantum communication (\Cref{prob:inner_product_k}) using $O(\med\{1/\epsilon^2,2^{n/2}/\epsilon,2^{n-k}/\epsilon^2\})$ copies of $\rho$ and $\sigma$.
\end{theorem}

\noindent Our protocol is a combination of the distributed inner product estimation protocol with single-copy measurements~\cite{anshu2022distributed}, which gives the upper bound of $O(\max\{1/\epsilon^2,2^{n/2}/\epsilon\})$, and a partial swap test algorithm (see \Cref{sec:partial_swap} for details) on the first $k$ qubits of $\rho$ and $\sigma$, which gives the upper bound of $O(2^{n-k}/\epsilon^2)$. In the regime where $\epsilon\geq1/\poly(n)$ is at least inverse polynomial in system size, the upper bound can be further reduced to $O(\min\{2^{n/2}/\epsilon,2^{n-k}/\epsilon^2\})$. Combining purity estimation with inner product estimation using the algorithm in \Cref{thm:partial_swap_informal}, we can estimate fidelity-like measures~\cite{elben2020cross,liang2019quantum} and the squared Hilbert-Schmidt distance $D_{\text{HS}}^2(\rho,\sigma)=\tr(\rho^2)+\tr(\sigma^2)-2\tr(\rho\sigma)$ with restricted quantum communication and memory. 

Given the algorithm in \Cref{thm:partial_swap_informal}, it is natural to ask about the optimality of the algorithm. Our next result provides a lower bound for \Cref{prob:purity_k} and thus a lower bound for \Cref{prob:inner_product_k}:
\begin{theorem}[Informal, see \Cref{thm:purity_ip_lower}]\label{thm:purity_ip_lower_informal}
For any $0\leq\epsilon<1$, any (possibly adaptive) protocols for estimating purity with $k$-qubit memory (inner product estimation with one-way $k$-qubit quantum communication) requires at least $\Omega(\med\{1/\epsilon^2,2^{n/2}/\sqrt{\epsilon},2^{n-k}/\epsilon^2\})$ copies of $\rho$ ($\rho$ and $\sigma$).
\end{theorem}

\noindent It is helpful to interpret this first in the regime where $\epsilon\geq 1/\poly(n)$. Within this regime for both purity estimation and inner product estimation, the sample complexity lower bound further reduces $\Omega(\min\{2^{n/2}/\sqrt{\epsilon},2^{n-k}/\epsilon^2\})$. Our result indicates that with $k\leq n/2$ qubits of quantum memory and quantum communication, no protocol can improve upon the sample complexity achieved by single-copy measurements and distributed inner product estimation with single-copy local measurements, respectively. This sample-memory tradeoff is similar to the sample-memory tradeoff for purity testing~\cite{ekert2002direct,chen2024optimal,chen2022exponential}, which can be solved by purity estimation with $\epsilon\simeq1/2$. However, the proof of sample complexity lower bound in \Cref{thm:purity_ip_lower_informal} is significantly different to obtain $\epsilon$ dependence. For general $\epsilon$, the lower bound, which can also be written as $\Omega(\max\{1/\epsilon^2,\min\{2^{n/2}/\sqrt{\epsilon},2^{n-k}/\epsilon^2\})$ can be understood approximately as a combination of the bound $\Omega(1/\epsilon^2)$ for tiny accuracy regime where $\epsilon=O(2^{-n})$, and $\Omega(\max\{2^{n-k}/\epsilon^2,2^{n/2}/\sqrt{\epsilon}\})$ for the tradeoff of estimation with $k$-qubit of quantum memory (one-way quantum communication).

We note that compared to \Cref{thm:partial_swap_informal} which gives an $O(\med\{1/\epsilon^2,2^{n/2}/\epsilon,2^{n-k}/\epsilon^2\})$ sample complexity upper bound, the lower bound in \Cref{thm:purity_ip_lower_informal} exhibits a $\sqrt{\epsilon}$ gap from the $2^{n/2}/\epsilon$ term in the upper bound. While for general quantum protocols closing this gap remains open, we prove an improved lower bound of $\Omega(2^{n/2}/\epsilon)$ for purity estimation (and thus distributed inner product estimation) using an identical projection-valued measurement (PVM) in every iteration:
\begin{theorem}[Informal, see \Cref{thm:improved_lower}]\label{thm:improved_lower_informal}
For any $0\leq\epsilon<1$, any protocols for estimating purity and distributed inner product estimation using an identical single-copy PVM in every round requires at least $\Omega(\max\{1/\epsilon^2,2^{n/2}/\epsilon\})$ copies of $\rho$.  
\end{theorem}
\noindent Although the assumption that the algorithm implements only one identical PVM throughout the protocol looks too restrictive, it captures the known algorithms on distributed inner product estimation and purity estimation~\cite{elben2020cross,anshu2022distributed}, and indicates that these protocols are optimal under this measurement restriction.

Besides the task of inner product estimation and purity estimation, we also investigate a decisional variants of quantum distributed inner product estimation, \emph{mixed state decisional inner product estimation}.
We obtain the strongest lower bounds for the following version:
\begin{problem}[DIPE: Convex mixture of Haar random states]\label{problem:DIPEconvexIntro}
Alice and Bob are given copies of states promised that one of the following two cases hold:
\begin{itemize}
    \item Alice and Bob both have state $\rho=\frac1{r}\sum_{i=1}^r |\psi_i\rangle\langle\psi_i|$, where the states $\psi_i$ are drawn iid from $\mu$ (the Haar measure).
    \item Alice has state $\rho=\frac1{r}\sum_{i=1}^r |\psi_i\rangle\langle\psi_i|$ and Bob has state $\sigma=\frac1{r}\sum_{i=1}^r |\phi_i\rangle\langle\phi_i|$ where the states $|\phi_i\rangle$ and $|\psi_i\rangle$ are drawn iid from $\mu$.
\end{itemize}
Their goal is to decide which case they are in with high probability using only LOCC.
\end{problem}
\noindent When $r=1$, and $\rho$ and $\sigma$ are pure states,~\Cref{problem:DIPEconvexIntro} reduces to \emph{decisional inner product estimation}, for which $\Theta(2^{n/2})$ copies are shown to be sufficient and necessary when Alice and Bob can only use LOCC~\cite{anshu2022distributed}. In \Cref{problem:DIPEconvexIntro}, the inner product between the states of Alice and Bob is $\mathbb{E}\tr(\rho^2)\sim 1/r$ in the first case, while in the second case, the inner product is approximately $2^{-n}$. Assuming $\tr(\rho^2)=\omega(2^{-n})$, we can solve it using a distributed quantum inner product estimation algorithm with accuracy $\tr(\rho^2)$, which uses $O(\max\{2^{n/2}/\tr(\rho^2)\})$ copies. 
We prove the following theorem in~\Cref{section:convexlowerbound}.
\begin{theorem}\label{thm:mix_dipe_lower_informal}
Any protocols for mixed state DIPE with fixed $\rho$ requires at least $\Omega(\sqrt{d/\mathbb{E}\tr(\rho^2)})$ copies even if Alice and Bob are allowed to perform arbitrary interactive protocols (or arbitrary LOCC operations).
\end{theorem}
Additionally we consider another version of mixed DIPE in Section~\ref{section:lowerboundfixed} where Alice and Bob either have the same state $U\rho U^{\dagger}$ or two independent states $U\rho U^{\dagger}$ and $V\rho V^{\dagger}$ for two independent Haar random unitaries $U$ and $V$.
For this variant we restrict to one-way classical communication between Alice and Bob, but we obtain a particularly direct proof using the concentration of measure phenomenon.

\subsection{Related works}
From a high-level perspective, our work is a part of the sequence of recent works exploring the relationship between near-term constraints on quantum devices and algorithms for various quantum learning tasks, and the underlying statistical complexity, see e.g.~\cite{aharonov2022quantum,bubeck2020entanglement,chen2021hierarchy,chen2022complexity,chen2022tight,chen2022tight2,chen2024tight,chen2022toward,fawzi2023quantum,fawzi2023lower,grewal2023efficient,huang2022quantum,huang2021information,liu2024quantum,liu2024role}. We refer to the survey~\cite{anshu2023survey} for a more thorough overview along this line of works and beyond.

\paragraph{Cross-platform verification.} The task of cross-platform verification was first proposed in Ref.~\cite{elben2020cross}, together with a protocol based on randomized measurements for comparing quantum states on different platforms. The same protocol is further analyzed from both theoretical and experimental perspectives in Refs~\cite{carrasco2021theoretical,greganti2021cross}, while a rigorous sample complexity bound for the key subroutine, distributed inner production estimation, is provided in Ref.~\cite{anshu2022distributed}. Recently, an efficient algorithm for inner product estimation based on Pauli sampling has been provided for quantum states with low magic and entanglement~\cite{hinsche2024efficient}. An experimental demonstration of this protocol among different existing quantum platforms is provided in Ref.~\cite{zhu2022cross}. An alternative cross-platform protocol using deep learning is proposed in Ref.~\cite{qian2023multimodal}. While all the above protocols allow only classical communication among different quantum platforms, Ref.~\cite{knorzer2023cross} proposes a protocol with limited quantum communication. A cross-platform comparison protocol between quantum processes is also developed by extending the original protocol for comparing quantum states~\cite{zheng2024cross}.

\paragraph{Testing and learning properties of quantum states.}
In this work, we can regard the task of testing properties of an unknown quantum system~\cite{aharonov2022quantum} as comparing Alice’s unknown quantum state to the classical description of a known quantum state. 
When we are comparing by computing the fidelity between the unknown quantum state and the description, direct fidelity estimation~\cite{flammia2011direct,da2011practical} provides efficient protocols based on joint measurements. 
We refer to Refs.~\cite{buadescu2019quantum,montanaro2013survey} for more general tasks of testing properties of quantum states. 
However, as shown by this and previous works~\cite{anshu2022distributed,hinsche2024efficient}, testing properties among multiple quantum systems in a distributed setting is very different from testing a single system. 

Quantum inner product estimation and purity estimation considered in this work also involve learning a specific property of Alice and Bob’s quantum systems. 
A most canonical task in learning properties of quantum states is to completely recover the density matrix to high accuracy, which is known as quantum state tomography~\cite{banaszek2013focus,blume2010optimal,gross2010quantum,hradil1997quantum}. 
Yet, the sample complexity for this task is proved to be exponential in the system size $n$~\cite{odonnell2016efficient,haah2016sample}. A well-studied approach that circumvents this exponential barrier is shadow tomography~\cite{aaronson2018shadow}, in which one only needs to approximate the expectation values of
$M$ observables of the unknown state to additive error $\epsilon$. Based on highly entangled joint measurements, a line of works achieve $\poly(\log M,n,1/\epsilon)$ sample complexity~\cite{aaronson2018online,aaronson2019gentle,buadescu2021improved,brandao2019quantum,chen2024optimalshadow,gong2023learning,watts2024quantum}. While the entangled measurements in the above protocols make them impractical to implement on near-term devices, Ref.~\cite{huang2020predicting} proposed an algorithm often referred to as classical shadows that uses $O(||O||_F\log M/\epsilon^2)$ single-copy measurements, where $||O||_F$ is the Frobenius norm of the estimated observables.
For general observables $O$ with $||O||_{\infty}\leq 1$, this will scale as $O(2^n\log M/\epsilon^2)$, which is shown to be optimal in Ref.~\cite{chen2022exponential}. 
When the observables are restricted to Pauli observables, an instance-optimal sample complexity is proposed for arbitrary subsets of Pauli observables and arbitrary measurement schemes~\cite{chen2024optimal}.

\paragraph{Quantum memory and communication tradeoffs.} 
Exploring the quantum memory tradeoffs for learning with and without bounded quantum memory has been widely investigated by a long line of recent works. 
A polynomial sample complexity gap for mixedness testing with and without quantum memory was first demonstrated in Ref.~\cite{bubeck2020entanglement}. 
Subsequent works~\cite{chen2022tight2,chen2022toward,aharonov2022quantum,chen2022exponential,huang2021information,huang2022quantum} further demonstrated exponential separations for a quantum memory for various tasks including shadow tomography, unitary learning, purity testing, and quantum dynamics learning. 
In particular, the relationship between the number of qubits in quantum memory and estimating Pauli observables is fully characterized in Refs~\cite{chen2022exponential,chen2024optimal}. Memory-sample tradeoffs are also revealed for other tasks such as purity testing~\cite{chen2024optimal} and Pauli channel eigenvalue estimation~\cite{chen2022quantum,chen2023efficientPauli}. 
In addition, a classical learning problem that exhibits a quantum memory trade-off was recently proposed~\cite{liu2023memory}. 

Similar intuition motivates a line of works regarding the sample (query) complexity tradeoffs for algorithms with and without bounded quantum communication. Over the past two decades, several works have established the separations between quantum over classical communication complexity in various settings, see e.g.~\cite{klauck2001interaction,klauck2005quantum}.

\paragraph{Concurrent work.} Arunachalam and Schatzki~\cite{arunachalam2024distributed} have related results concurrent to our work and we thank them for the helpful exchange. In particular, they investigate the distributed property testing setting where Alice and Bob are to estimate $\bra{\psi}M\ket{\phi}$ given $\ket{\psi}$ and $\ket{\phi}$, respectively. They also give a tight bound with respect to $n$ and $k$ when Alice and Bob are to estimate $\abs{\braket{\psi|\phi}}^2$ with $k$-qubit of quantum communication in total.

\subsection{Outlook}
In this work, we propose protocols for quantum inner product estimation with limited quantum communication and unentangled measurements and quantum purity estimation with limited quantum memory, and prove that they are near-optimal to at most a $\sqrt{\epsilon}$ factor. Several open problems remain to be answered:

\paragraph{Arbitrary $k$-qubit communication.} In \Cref{prob:inner_product_k}, we consider quantum inner product estimation when Alice and Bob have $k$-qubit one-way communication and one-way classical communication. It is thus natural to ask if the lower bound in this work holds for the case when Alice and Bob have $k$-qubit arbitrary quantum communication and arbitrary LOCC.

\paragraph{Tight bounds.} While our protocols yield sample complexity upper bound $O(\med\{1/\epsilon^2, 2^{n/2}/\epsilon, 2^{n-k}/\epsilon^2\})$ for inner product estimation with $k$-qubit quantum communication and unentangled measurements and quantum purity estimation with $k$-qubit quantum memory, there is a $\sqrt{\epsilon}$ gap from the lower bound of $\Omega(\med\{1/\epsilon^2, 2^{n/2}/\sqrt{\epsilon}, 2^{n-k}/\epsilon^2\})$ we obtained in this work. A similar $\sqrt{\tr(\rho^2)}$ gap appears for the mixed state DIPE problem. An immediate open question is to find the true $\epsilon$ and $\tr(\rho^2)$ dependence for the sample complexity of these two tasks.

\paragraph{Alternative models for restricted quantum communication.}We consider the setting in which Alice and Bob are only allowed to perform unentangled local measurements and share $k$ qubits of quantum communication for each pair of $\rho$ and $\sigma$. An alternative model for restricted quantum communication is when Alice and Bob can perform local joint measurements but can only send $k$ qubits of quantum communication in total. It is interesting to see the sample complexity under this model.

\paragraph{Other distributed quantum property estimation problems.}It is an interesting future direction to explore if our techniques can yield protocols or lower bounds for other distributed quantum property estimation problems of the form $\tr(\rho\otimes\sigma\cdot W)$ for some two-body operators $W$ with only classical communication or restricted quantum communication.

\subsection{Roadmap}
In \Cref{sec:prelim} we provide technical preliminaries. In \Cref{sec:purity_to_ip}, we compare the tasks of inner product estimation with $k$-qubit one-way quantum communication and purity estimation with $k$-qubit quantum memory, and show the connection between the sample complexity between these models. We prove the lower bounds for purity estimation in \Cref{thm:purity_ip_lower_informal} and \Cref{thm:improved_lower_informal} in \Cref{sec:lower_memory} and \Cref{sec:improved_lower}. We prove the upper bound for inner product estimation with $k$-qubit quantum communication (\Cref{thm:partial_swap_informal}) in \Cref{sec:partial_swap}. Finally, we prove the lower bound for mixed state DIPE in \Cref{thm:mix_dipe_lower_informal} in \Cref{sec:mix_dipe}.

\section{Preliminaries}\label{sec:prelim}

In this section, we provide the basic concepts and results required throughout this paper. We use $\norm{A}$, $\norm{A}_2$, and $\norm{A}_1$ to represent the operator norm, Euclidean norm, and trace norm of matrix $A$. For vectors, we use $\norm{v}_p$ to denote the $L_p$ norm of vector $v$. The total variation distance between probability distributions $p$ and $q$ are denoted as $\norm{p-q}_{\text{TV}}$. We will abbreviate $\{1, 2, \cdots, m\}$ as $[m]$, $(x_{i_1}, x_{i_1+1}, \cdots, x_{i_2})$ as $x_{[i_1:i_2]}$. We use the standard big-O notation $(O,\Omega)$ throughout. When we say ``with high probability'' without specification, we mean with probability at least $2/3$. We denote the $n$-qubit identity operator by $I_n$ and ignore $n$ when it is clear from the context.

\subsection{Basic results in quantum information}\label{sec:quantum_info}
We start with the standard definitions and folklore in quantum information. In general, a $n$-qubit quantum state can be represented as a positive semi-definite matrix $\rho\in\mathbb{C}^{2^n\times 2^n}$ with $\tr(\rho)=1$. In the following, we also use $d=2^n$ to denote the dimension of a $n$-qubit quantum state. A pure state, which is usually represented as $\ket{\psi}$, $\ket{\phi}$, or $\ket{\varphi}$ throughout this paper, is a rank-$1$ state and thus $\tr(\rho^2)=1$. Given a (possibly unnormalized) $n$-qubit quantum state $\rho$ and a subset $S\subseteq [n]$, we use the partial trace $\tr_{S}(\rho)$ to denote the remaining state after tracing out the qubits in $S$. %The Bell states (bases) $\{\ket{\Psi^\pm}=(\ket{10}\pm\ket{01})/\sqrt{2},\ket{\Phi^\pm}=(\ket{00}\pm\ket{11})/\sqrt{2}\}$ is a set of maximally entangled two-qubit pure states. 
A $\SWAP_n$ gate is a $2n$-qubit operator that swaps the first and last $n$ qubits. We omit $n$ when it is clear from the context.

\paragraph{Quantum measurements.}Quantum measurements are represented as positive operator-valued measures (POVMs) in the general case. An $n$-qubit POVM is defined by a set of positive-semidefinite matrices $\{F_s\}_s$ with $\sum_s F_s=I$, where each $F_s$ is a \emph{POVM element} corresponding to a \emph{measurement outcome} $s$. The probability of observing outcome $s$ when we measure a given quantum state $\rho$ using the POVM $\{F_s\}_s$ is $\tr(F_s\rho)$. 

When $F_s$ is rank-$1$ for all $s$, we call $\{F_s\}_s$ a \emph{rank-$1$ POVM}. Lemma 4.8 in Ref.~\cite{chen2022exponential} proves that any POVM can be simulated by a rank-$1$ POVM and classical post-processing information-theoretically. We will thus assume when we prove our lower bounds in the following parts of the paper that all measurements in the learning protocols only use rank-$1$ POVMs. We can thus parametrize any rank-$1$ POVMs as
\begin{align*}
    \{w_s2^n\ket{\psi_s}\bra{\psi_s}\}\,,
\end{align*}
for pure states $\{\ket{\psi_s}\}$ and non-negative weights $w_s$ with $\sum_s w_s = 1$. When $w_s=2^{-n}$ for all $s$, the POVM elements satisfies $F_s=\ket{\psi_s}\bra{\psi_s}$ and the POVM then reduces to a projection-valued measurement (PVM).

We also consider the post-measurement quantum state after the POVM $\{F_s\}$. The post-measurement state is not determined by the POVM $\{F_s\}_s$ itself, but by the physical realization of it. In particular, we need to specify a decomposition for every POVM element $F_s$ in $\{F_s=M_s^\dagger M_s\}_s$, where $M_s$ is a $2^m\times 2^n$ matrix. The post-measurement state of $\rho$ corresponding to the outcome $s$ is an $m$-qubit state given by
\begin{align*}
    \rho\to\frac{M_s\rho M_s^\dagger}{\tr(M_s^\dagger M_s\rho)}.
\end{align*}
We will refer to such a (realization of a) POVM as having \emph{$n$-qubit input and $m$-qubit output} or simply as being an \emph{$(n\to m)$-POVM} to emphasize the dimension of the post-measurement state. We say ``POVM'' in this paper for a realization with $0$-qubit output unless specify the number of qubits in the output. 

\paragraph{Measurements with classical communications} In this paper, we usually have to consider the measurement $\{F_s\}_s=\{M_{A,B}^s\}_s$ that is jointly performed by two parties Alice and Bob. We ignore the outcome $s$ when its does not matter in the context. Suppose Alice and Bob can only have classical communications. If the classical is restricted to one-way communication, we can always depict the measurement scheme for a two-case distinguishing tasks as follow: Alice performs a POVM $\{M_i\}$ on her copies and sends the result $i$ to Bob. After receiving $i$, Bob performs a two-outcome POVM $\{N_i,I-N_i\}$ on his copies to decide which case they are in. We thus have every $M_{A,B}$ is in the form of
\begin{align}
M_{A,B}=\sum_i M_i\otimes N_i,
\end{align}
with $\sum_i M_i=I$.

If Alice and Bob can have arbitrary classical communication, $\{M_{A,B}^s\}_s$ can be described as quantum measurements with local quantum operations and classical communication (LOCC). A rigorous mathematical representation for measurements with LOCC is complicated. Instead we consider separable measurements, which is defined as follows
\begin{definition}[Separable measurements]\label{def:separable_measurements}
nts). A two-outcome POVM $\{M_{A,B}, I-M_{A,B}\}$ is separable if and only if
\begin{align*}
M_{A,B}=\sum_t A_t\otimes B_t,\quad I-M_{A,B}=\sum_t A'_t\otimes B'_t,\quad A_t, A'_t,B_t,B'_t\geq 0.
\end{align*}
\end{definition}
It is proved that any quantum measurements with LOCC is included in the class of separable measurements~\cite{bennett1999quantum,chitambar2009nonlocal}.

\subsection{A brief review of Haar random unitaries}\label{sec:rand_unitary}
A common tool to prove lower bounds on the sample complexity of learning problems is to consider random instances.
The random states we consider here are based on Haar random unitaries.
The Haar measure $\mu$ on the unitary group $U(d)$ is the unique probability measure that is invariant under left- and right-multiplication.
That is, 
\begin{equation*}
\underset{U\sim\mu}{\mathbb{E}}f(UV)=\underset{U\sim\mu}{\mathbb{E}} f(VU)=\underset{U\sim\mu}{\mathbb{E}}  f(U)
\end{equation*}
for any unitary $V\in U(d)$.
We can also define a unique rotation invariant measure on states by $U|\psi\rangle$ with $U\sim \mu$.
Here, $|\psi\rangle\in \mathbb{C}^d$ can be any quantum state.
Slightly abusing notation we will write $\psi\sim \mu$.

The Haar measure has multiple desirable properties for proving lower bounds, most noticeably \textit{concentration of measure} in the form of Levy's lemma:
A function $f:U(d)\to \mathbb{R}$ is called Lipshitz-continuous with Lipshitz constant $L$ if 
\begin{equation*}
    |f(U)-f(V)|\leq L ||U-V||_{2},
\end{equation*}
where $||\bullet||_2$ denotes the Frobenius or Schatten 2-norm.
\begin{theorem}[Levy's lemma for Haar-random pure states] \label{thm:levy}
	Let $f: U(d) \to \mathbb{R}$ be a function  that is Lipschitz with Lipschitz constant $L$. Then,
	\begin{equation*}
		\underset{U \sim \mu} {\mathrm{Pr}}\left[\left| f(U) - \underset{U\sim\mu}{\mathbb{E}}\left[ f(U)\right] \right| >\tau \right] \leq 4 \exp \left( - \frac{2d\tau^2}{9 \pi^3 L^2} \right) \quad \text{for any $\tau>0$.}
	\end{equation*}
\end{theorem}
We will apply this to the outcome probabilities of measurements using $k$-copies of $U\sim \mu$.
We can then use Theorem~\ref{thm:levy} to show that, for small $k$, the outcomes of such measurements do not depend on $U$ and therefore reveal only little information.

In some settings we will need to explicitly compute expectation values over the Haar measure.
A key tool is the following folklore formula:
\begin{lemma}[See e.g. Ref.~\cite{harrow2013church}]\label{lem:Haar_avg}
\begin{equation*}
    \underset{\psi\sim \mu}{\mathbb E}(|\psi\rangle\langle \psi|)^{\otimes k}=\frac{\mathcal{S}_k}{{{d+k-1} \choose k}}=\frac{1}{d\cdots (d+k-1)}\sum_{\pi\in S_k} \pi^d,
\end{equation*}
where $\mathcal{S}_k$ denotes the projector onto the symmetric subspace $\mathrm{Sym}^k(\mathbb{C}^d)$, $S_k$ is the set of permutation over $k$ elements and $\pi^d$ acts on $(\mathbb{C}^d)^{\otimes k}$ via
\begin{equation*}
    \pi^d |i_1,\ldots,i_k\rangle=|i_{\pi^{-1}(1)},\ldots, i_{\pi^{-1}(k)}\rangle.
\end{equation*}
\end{lemma}
% We note that in our paper, for permutation $\pi$, only $\pi^{-1}$ denotes the inverse of $\pi$, while $\pi^D,\pi^d,\pi^{1/\epsilon}$ denote the linear transformation defined above.

\subsection{Tree representations and the Le Cam’s two-point method}\label{sec:tree}
In this paper we focus on adaptive protocols, which can be naturally modeled as learning trees \cite{aharonov2022quantum, bubeck2020entanglement, chen2022exponential, huang2022provably, chen2024optimal}. In each node, we select $c$ copies of the unknown state $\rho$, perform some POVM on $\rho^{\otimes c}$, and step to its child node corresponding to the outcome. Adapting Ref.~\cite{chen2024optimal}, we call such protocol a $(c, \cM)$ learning protocol, where $\cM$ is the set of feasible POVMs on $\rho^{\otimes c}$. Throughout the paper, we only consider two specific kinds of protocols: 
\begin{itemize}
    \item $c=1$ and $\cM$ is the set of all single-copy POVMs. This characterizes all protocols that only use single-copy measurements.
    \item $c=2$ and $\cM$ is the set of all rank-$1$ two-copy POVMs where there are only $k$ qubits of quantum memory (See~\Cref{def:learning_k} for the formal definition).
\end{itemize}

\begin{definition}[Tree representation for $(c, \cM)$ learning protocols \cite{chen2024optimal}]\label{def:TreeRepNonCM}
    Given an unknown $n$-qubit quantum state $\rho$, a $(c, \cM)$ learning protocol can be represented as a rooted tree $\cT$ of depth $T$ with each node on the tree recording the measurement outcomes of the algorithm so far. In particular, it has the following properties:
    \begin{itemize}
        \item We assign a probability $p^\rho(u)$ for any node $u$ on the tree $\cT$.
        \item The probability assigned to the root $r$ of the tree is $p^\rho(r)=1$.
        \item At each non-leaf node $u$, we measure $\rho^{\otimes c}$ using an adaptively chosen POVM $M_u=\{F_{s}^u\}_s\in \cM$, which results in a classical outcome $s$. Each child node $v$ corresponding to the classical outcome $s$ of the node $u$ is connected through the edge $e_{u,s}$.
        \item If $v$ is the child of $u$ through the edge $e_{u,s}$, the probability to traverse this edge is $p^{\rho}(s|u)= \tr(F_s^u\rho^{\otimes c})$. Then
        \begin{align*}
        p^\rho(v)=p^\rho(u)p^\rho(s|u)=p^\rho(u)\tr(F_s^u \rho^{\otimes c}).
        \end{align*}
        \item Each root-to-leaf path is of length $T$. Note that for a leaf node $\ell$, $p^\rho(\ell)$ is the probability for the classical memory to be in state $\ell$ at the end of the learning protocol. Denote the set of leaves of $\cT$ by $\mathrm{leaf}(\cT)$
    \end{itemize}
    At the end of the protocol, a classical algorithm $\cA$ maps any leaf node $\ell$ to a desired output of the protocol. $\cA$ is called the post-processing part of the protocol. 
\end{definition}

To prove the lower bound of purity estimation, we consider the reduction from the following distinguishing problem:

\begin{problem}[Distinguishing the maximally mixed state and an ensemble $\cD$]\label{prob: distinguishing_general}
    Given access to copies of an unknown state $\rho$ of dimension $d=2^n$, distinguish the following two cases: %with probability $\frac{7}{12}$:
    \begin{itemize}
        \item $\rho$ is the maximally mixed state $\rho_m=I/2^n$.
        \item $\rho$ is sampled from a known distribution $\cD$ over $n$-qubit mixed states.
    \end{itemize}
\end{problem}

We will explicitly construct an ensemble $\cD_{n, \epsilon}$ (\Cref{def: random induced states}) such that the distinguishing problem can be reduced to purity test. There are mature tools for proving lower bound of a many-versus-one distinguishing problem, including the Le Cam's two-point method~\cite{yu1997assouad}, one-sided likelihood ratio~\cite{chen2022exponential}, and the martingale technique~\cite{chen2022complexity}. Here we adapt notations and statements in Ref.~\cite{chen2024optimal}.

\begin{definition}[Likelihood ratio]
    Let $\cT$ be the tree representation of a learning protocol for the many-versus-one distinguishing task in \Cref{prob: distinguishing_general}. For any $\ell\in \mathrm{leaf}(\cT)$, define the likelihood ratio 
    \[L(\ell)\coloneqq \frac{\E_{\psi\sim \cD}p^{\psi}(\ell)}{p^{\rho_m}(\ell)}.\]
    We also define the likelihood ratio for each edge $e_{u,s}$ and each state $\psi$:
    \[L_\psi(u, s)\coloneqq \frac{p^{\psi}(s|u)}{p^{\rho_m}(s|u)},\quad L_\psi(\ell)\coloneqq \frac{p^{\psi}(\ell)}{p^{\rho_m}(\ell)}.\]
\end{definition}

\begin{lemma}[Toolbox for proving lower bounds]\label{lem: toolbox for lower bounds}
    Suppose $\cT$ is a learning tree with depth $T$ that solves the distinguishing problem in \Cref{prob: distinguishing_general} with probability $p_{\text{succ}}$.
    \begin{enumerate}[label=(\alph*)]
        \item (Le Cam's two-point method~\cite{yu1997assouad})
        \[p_{\text{succ}}\leq d_{TV}(\E_{\psi\sim \cD}p^\psi, p^{\rho_m})\coloneqq \frac12\sum_{\ell\in \mathrm{leaf}(\cT)}\abs{\E_{\psi\sim \cD}p^\psi(\ell)- p^{\rho_m}(\ell)}.\]
        \item (One-sided likelihood ratio~\cite{chen2022exponential}, see also \cite[Lemma 6]{chen2024optimal}) For any $\beta > 0$,
        \begin{equation*}
            d_{TV}(\E_{\psi\sim \cD}p^\psi, p^{\rho_m})\leq \Pr_{\ell\sim p^{\rho_m}(\ell), \psi\sim \cD}[L_\psi(\ell)\leq \beta] + 1-\beta,
        \end{equation*}
        \begin{equation*}
            d_{TV}(\E_{\psi\sim \cD}p^\psi, p^{\rho_m})\leq \Pr_{\ell\sim p^{\rho_m}(\ell)}[L(\ell)\leq \beta] + 1-\beta.
        \end{equation*}
        \item (Martingale technique~\cite{chen2022complexity}, see also \cite[Lemma 7]{chen2024optimal}). If there exists is a $\delta>0$ such that for every node $u$ we have 
        \begin{equation*}
            \E_{\psi\sim \cD}\E_{s\sim p^{\rho_m}(s|u)}[(L_\psi(u, s)-1)^2]\leq \delta,
        \end{equation*}
        then 
        \begin{equation*}
             \Pr_{\ell\sim p^{\rho_m}(\ell), \psi\sim \cD}[L_\psi(\ell)\leq 0.9]\leq 0.1 + c\delta T,
        \end{equation*}
        where $c$ is a universal constant.
    \end{enumerate}
\end{lemma}
These tools upper bound the success probability from different perspectives. \Cref{lem: toolbox for lower bounds}(a) is tree-based: we upper bound the statistical distance between the leaf distributions under the two cases. \Cref{lem: toolbox for lower bounds}(b) is path-based: we have to prove that for most of the paths from root to leaves, the likelihood ratios are not too small. \Cref{lem: toolbox for lower bounds}(c) is edge-based: we only have to focus on each edge and show that the likelihood ratio over the edge concentrates around 1.

\section{From purity to inner product estimation}\label{sec:purity_to_ip}
\begin{figure}[htbp]
    \centering
    \includegraphics[width=0.8\textwidth]{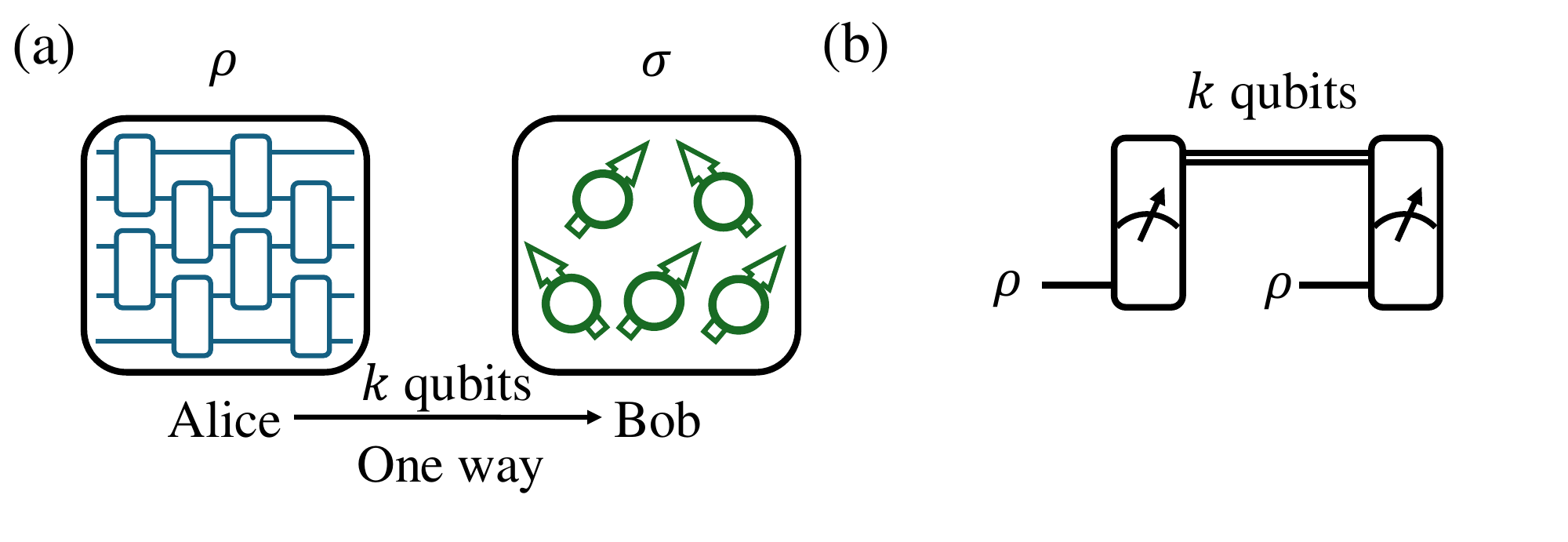}
    \caption{Comparison of the models for quantum inner product estimation with $k$-qubit one-way quantum communication and any classical one-way communication (\Cref{prob:inner_product_k}) and purity estimation with $k$-qubit quantum memory (\Cref{prob:purity_k}).}
    \label{fig:illustration}
\end{figure}

In this section, we consider the models for quantum inner product estimation with $k$-qubit one-way quantum communication (\Cref{prob:inner_product_k}) and purity estimation with $k$-qubit quantum memory (\Cref{prob:purity_k}), and show the connection between the sample complexity two tasks (see \Cref{fig:illustration} for illustration). 

For quantum inner product estimation defined in \Cref{prob:inner_product_k},  Alice and Bob are allowed to perform $k$-qubit one-way quantum communication, any one-way classical communication, and unentangled local measurements. As shown in \Cref{fig:illustration} (a), we can always describe such a protocol sequentially as the quantum communication is one-way:

\begin{definition}[Sequential description of \Cref{prob:inner_product_k}]\label{def:seq_inner_product}
The sequential description of a protocol for \Cref{prob:inner_product_k}
\begin{itemize}
    \item Alice takes a copy of $\rho$ and selects an $(n\to k)$-POVM $\{M_s^\dagger M_s\}_s$ (see \Cref{sec:quantum_info} for definitions) to measure $\rho$.  
    \item Alice obtains the classical outcome $s$ with probability $\tr(\rho\cdot M_s^\dagger M_s)$, send arbitrary classical message (might depend on $s$), and send the $k$-qubit post-measurement $\varrho_{s,k}=M_s\rho M_s^\dagger/\tr(\rho\cdot M_s^\dagger M_s)$ to Bob.
    \item Bob receives the classical message and $\varrho_{s,k}$, and performs an $(n+k\to 0)$-POVM $\{M_{s'}^\dagger M_{s'}\}_{s'}$ on $\varrho_{s,k}\otimes\sigma$.
\end{itemize}
\end{definition}

For purity estimation with $k$-qubit quantum memory defined in \Cref{prob:purity_k} (as shown in \Cref{fig:illustration} (b)), we can also provide a sequential definition as below adapting Ref.~\cite{chen2024optimal}:

\begin{definition}[Purity estimation with $k$ qubits of quantum memory, see also Definition 4 of Ref.~\cite{chen2024optimal}]\label{def:learning_k}
The algorithm consists of $N/2$ rounds and each round involves $2$ copies of $\rho$. In each round, the algorithm maintains a $k$-qubit quantum memory $\varrho$ (initialized to a fixed quantum state at first). Within the round, we get a new copy of $\rho$, select an $(n\to k)$ POVM $\{M_s^\dagger M_s\}_s$ to measure $\rho$, obtain the outcome $s$ with probability $\tr(\rho\cdot M_s^\dagger M_s)$, and store the $k$-qubit post-measurement state $\varrho_{s,k}=M_s\rho M_s^\dagger/\tr(\rho M_s^\dagger M_s)$ into the quantum memory. The algorithm then takes a new copy of $\rho$, and select an $(n+k\to 0)$ POVM $\{M_{s'}^\dagger M_{s'}\}_{s'}$ to measure $\varrho_{s,k}\otimes\rho$, obtain the outcome $s'$ with probability $\tr((\varrho_{s,k}\otimes\rho)\cdot M_{s'}^\dagger M_{s'})$. Finally, the quantum memory is reset to empty, and the next round begins. After $N/2$ rounds, the algorithm predicts purity $\tr(\rho^2)$ of $\rho$. The total sample complexity of the protocol is $N$.
\end{definition}
\noindent Under this learning model with $k$ qubit of quantum memory, purity testing, which distinguishes between $\rho$ being a Haar random pure state from the maximal mixed state, requires $\Theta(\min\{2^{n/2},2^{n-k}\})$ copies of $\rho$ for algorithms with $k$-qubit quantum memory~\cite{chen2024optimal}. However, for purity estimation to additive error $\epsilon$, the dependence on $\epsilon$ for sample complexity is unclear.

Given the sequential description of quantum inner product estimation with bounded one-way quantum communication in \Cref{prob:inner_product_k} and purity estimation with bounded memory in \Cref{def:learning_k} (\Cref{prob:purity_k}), we can observe the following connection between the sample complexity of \Cref{prob:inner_product_k} and \Cref{prob:purity_k}:

\begin{claim}\label{lem:connection}
Consider quantum inner product estimation with $k$-qubit one-way quantum communication (\Cref{prob:inner_product_k}) and purity estimation with $k$-qubit quantum memory (\Cref{prob:purity_k}). Suppose we have an algorithm $\mathcal{A}$ that can solve \Cref{prob:inner_product_k} with $N$ copies of $\rho$ and $\sigma$, we have an algorithm $\mathcal{A}'$ that solve \Cref{prob:purity_k} using $2N$ copies of $\rho$.
\end{claim}

\noindent The intuition for this claim is straightforward given the sequential descriptions of two tasks. Suppose we have an algorithm $\mathcal{A}$ for \Cref{prob:inner_product_k} using $N$ copies of $\rho$ and $\sigma$ (i.e., $N$ rounds in \Cref{def:seq_inner_product}), we can describe the algorithm $\mathcal{A'}$ in a similar pattern for \Cref{prob:purity_k}:
\begin{itemize}
    \item We take a copy of $\rho$ and select the same $(n\to k)$-POVM $\{M_s^\dagger M_s\}_s$ in $\mathcal{A}$ to measure $\rho$. 
    \item We obtain the classical outcome $s$ with probability $\tr(\rho\cdot M_s^\dagger M)$ and store the $k$-qubit post-measurement $\varrho_{s,k}=M_s\rho M_s^\dagger/\tr(\rho\cdot M_s^\dagger M)$ into the quantum memory.
    \item We take a new copy of $\rho$ and perform the same $(n+k\to 0)$-POVM $\{M_{s'}^\dagger M_{s'}\}_{s'}$ in $\mathcal{A}$ on $\varrho_{s,k}\otimes\rho$.
\end{itemize}
One can verify that the above sequence satisfies \Cref{def:learning_k} and $\mathcal{A}'$ is a protocol for \Cref{prob:purity_k} using $2N$ copies of $\rho$.

\section{Lower bounds for purity estimation}

\subsection{Universal lower bound for purity estimation}\label{sec:lower_entangled}
In this section, we prove a lower bound for purity estimation when arbitrary entangled measurements are allowed.
Since entangled measurements are the most general form of quantum algorithms, the lower bound also applies to the restrictive settings that we will consider later.

\begin{lemma}\label{lem:entangled_lower}
Any algorithm $\cA$ that estimates the purity within additive error $\epsilon\leq1/36$ using entangled measurements has sample complexity at least $\Omega(1/\epsilon^2)$.
\end{lemma}

\begin{proof}
Let $\delta=\frac{1}{6}-\sqrt{\frac{1}{36}-\epsilon}$.
Consider $\rho_0=\frac{1}{3}|0\rangle\langle0|+\frac{2}{3}|1\rangle\langle1|$ and $\rho_1=(\frac{1}{3}+\delta)|0\rangle\langle0|+(\frac{2}{3}-\delta)|1\rangle\langle1|$.
We have $\tr(\rho_0^2)=\frac{5}{9}$ and $\tr(\rho_1^2)=\frac{5}{9}-\frac{2}{3}\delta+2\delta^2$, so the purity of the two states differ by $\frac{2}{3}\delta-2\delta^2=2\epsilon$.
Thus we can use $\mathcal{A}$ to distinguish the two states based on whether the estimated purity is larger than $\frac{5}{9}-\epsilon$.
However, by Helstrom theorem~\cite{helstrom1969quantum}, the probability of distinguishing $t$ copies of $\rho_0$ and $t$ copies of $\rho_1$ is at most $\frac{1}{2}+\frac{1}{4}\|\rho_0^{\otimes t},\rho_1^{\otimes t}\|_1$.
Therefore $\|\rho_0^{\otimes t}-\rho_1^{\otimes t}\|_1\geq\frac{2}{3}$.
Hence, we have both
\begin{align*}
&F(\rho_0^{\otimes t},\rho_1^{\otimes t})\leq1-\frac{1}{4}\|\rho_0^{\otimes t}-\rho_1^{\otimes t}\|_2^2\leq\frac{8}{9},\ \text{and}\\ 
&F(\rho_0^{\otimes t},\rho_1^{\otimes t})=F(\rho_0,\rho_1)^t=\left(\sqrt{\frac{1}{3}\left(\frac{1}{3}+\delta\right)}+\sqrt{\frac{2}{3}\left(\frac{2}{3}-\delta\right)}\right)^{2t}\geq(1-40\epsilon^2)^t\geq1-40\epsilon^2t,
\end{align*}
where the second inequality uses the well-known relation that fidelity is upper bounded by $1$ minus the squared trace distance.
Thus the number of copies $t$ needed for $\cA$ is $\Omega(1/\epsilon^2)$.
\end{proof}

\subsection{Lower bound for purity estimation with bounded quantum memory}\label{sec:lower_memory}

In this section, we prove a lower bound of $\Omega(\min\{\sqrt{d/\epsilon},2^{n-k}/\epsilon^2\})$ for protocols with $k$ qubits of memory.
This together with the $\Omega(1/\epsilon^2)$ lower bound from \Cref{sec:lower_entangled} yields a lower bound of
\[\Omega(\max\{1/\epsilon^2,\min\{\sqrt{d/\epsilon},2^{n-k}/\epsilon^2\}\})=\Omega(\med\{1/\epsilon^2, 2^{n/2}/\sqrt{\epsilon}, 2^{n-k}/\epsilon^2\}).\]
When $k=0$, the protocols considered degenerate to protocols that only use single copy measurements.
This implies a $\Omega(\max\{1/\epsilon^2,2^{n/2}/\sqrt{\epsilon}\})$ lower bound for single copy measurements.

Our lower bound is proved via a reduction from \Cref{prob: distinguishing_general}. To this end, we need an appropriate ensemble $\cD$ of $n$-qubit mixed states such that: 
\begin{itemize}
    \item With high probability, a state $\psi$ sample from $\cD$ has high fidelity, i.e., $\tr(\psi^2)=\tr(\rho_m^2)+\Theta(\epsilon)$, so that an algorithm for purity estimation implies an algorithm for distinguishing $\rho_m$ and $\cD$.
    \item There is a neat expression of $\E_{\psi\sim \cD}\psi^{\otimes T}$ that allows us to prove lower bound using tools from \Cref{lem: toolbox for lower bounds}.
\end{itemize}

We achieve the two conditions by tracing out part of qubits from a large Haar random state. This is called a random induced state~\cite{bengtsson2017geometry} which has been considered extensively in the literature of quantum entanglement~\cite{aubrun2012partial,hayden2006aspects,aubrun2014entanglement}.
To the best of our knowledge, this is the first time it is considered for quantum hypothesis testing.

\begin{definition}[Ensemble of random induced states]\label{def: random induced states}
    Define $\cD_{n, \epsilon}$ as the distribution of $\psi=\tr_{>n}(\ketbra{H})$, where $\ket{H}$ is a $D=2^{n+\log(1/\epsilon)}$ dimensional Haar random state, and $\tr_{>n}$ means tracing out the last $\log(1/\epsilon)$ qubits.
\end{definition}

We now instantiate \Cref{prob: distinguishing_general} into the following problem and show that the problem can be reduced to purity estimation.
\begin{problem}\label{prob: distinguishing}
    Distinguish the maximally mixed state $\rho$ and the ensemble $\cD_{n, \epsilon}$ of random induced states (in the sense of \Cref{prob: distinguishing_general}) with probability at least $7/12$.
\end{problem}

% An algorithm for purity estimation immediately implies an algorithm for \Cref{prob: distinguishing} since with high probability $\tr(\psi^2)>\tr(\rho_m^2)+\epsilon$ (see \Cref{lem:purity_to_dist}). Therefore, we only need to prove lower bound for \Cref{prob: distinguishing}. 

\begin{lemma}\label{lem:purity_to_dist}
An algorithm $\cA$ that estimates purity within additive error $\epsilon/3$ can be used to construct an algorithm that solves \Cref{prob: distinguishing} with the same samples complexity.
\end{lemma}

\begin{proof}
We need some basic properties of random induced states, which can be obtained from direct calculations using Lemma~\ref{lem:Haar_avg}:
\[\E_{\psi\sim\cD_{n,\epsilon}}\tr(\psi^2)=\frac{2^n\epsilon+1}{2^n+\epsilon}\]
\[\Var_{\psi\sim\cD_{n,\epsilon}}\tr(\psi^2)=\frac{2(2^{2n}-1)(1/\epsilon^2-1)}{(2^n/\epsilon+1)^2(2^n/\epsilon+2)(2^n/\epsilon+3)}.\]
By Chebyshev's inequality,
\[p=\Pr_{\psi\sim\cD_{n,\epsilon}}\Bigl[\tr(\psi^2)\leq2^{-n}+\frac{2\epsilon}{3}\Bigr]\leq\frac{\Var_{\psi\sim\cD_{n,\epsilon}}\tr(\psi^2)}{(\E_{\psi\sim\cD_{n,\epsilon}}\tr(\psi^2)-(2^{-n}+2\epsilon/3))^2}=O(2^{-2n}).\]
To solve \Cref{prob: distinguishing}, use $\cA$ to estimate the purity and output $\rho_m$ if the estimated purity is smaller than $2^{-n}+\epsilon/3$.
With probability $1-p$, $\rho$ in the random induced state case has purity greater than $2^{-n}+2\epsilon/3$.
Then, with probability at least $\frac{2}{3}$ $\cA$ estimates purity within error $\epsilon/3$, in which case the distinguishing task in \Cref{prob: distinguishing} is solved correctly.
Thus the probability of solving the distinguishing task correctly is at least $(1-p)2/3\geq\frac{7}{12}$.
\end{proof}

Hence it suffices to prove lower bounds for \Cref{prob: distinguishing}.
The techniques are similar to the ones used in Ref.~\cite{chen2024optimal}.
We include some lemmas there:

\begin{lemma}[See Lemma 16 of Ref.~\cite{chen2024optimal}]\label{lem:perm_ineq}
Let $x,y$ be two positive integers, $\rho_x$ be a mixed state on $x$ qudits (each qudit has local dimension $d$), and $\rho_y$ be a mixed state on $y$ qudits. Then
\[\tr\Bigl(\rho_x\otimes \rho_y\sum_{\pi\in S_{x+y}}\pi^d\Bigr)\geq\tr\Bigl(\rho_x\sum_{\pi\in S_x}\pi^d\Bigr)\tr\Bigl(\rho_y\sum_{\pi\in S_y}\pi^d\Bigr).\]
\end{lemma}

\begin{lemma}[See Lemma 17 of Ref.~\cite{chen2024optimal}]
\[\max_{\{F_s\}_s\in\cM_{2,n}^k}\sum_s\frac{\tr(F_s\SWAP)^2}{\tr(F_s)}\leq2^{k+n}.\]
\end{lemma}

Now we are ready for the lower bound.
\begin{theorem}\label{thm:purity_ip_lower}
Any algorithm $\cA$ that estimates the purity within additive error $\epsilon$ with $2$-copy protocols with $k$ qubits of quantum memory requires $\Omega(\med\{1/\epsilon^2, 2^{n/2}/\sqrt{\epsilon}, 2^{n-k}/\epsilon^2\})$ copies.
\end{theorem}
\noindent By \Cref{lem:connection}, \Cref{thm:purity_ip_lower} also indicates a $\Omega(\med\{1/\epsilon^2, 2^{n/2}/\sqrt{\epsilon}, 2^{n-k}/\epsilon^2\})$ sample complexity for quantum inner product estimation using $k$-qubit one-way communication as in \Cref{prob:inner_product_k}.

\begin{proof}
As argued before, it suffices to prove a $\Omega(\min\{2^{n/2}/\sqrt{\epsilon},2^{n-k}/\epsilon^2\})$ lower bound for \Cref{prob: distinguishing}.
Using the tree representation, consider a path $\{e_{u_t,s_t}\}_{t=1}^T$ from root leaf $\ell$.
We have the likelihood ratio for $\psi$
\[L_\psi(\ell)=\prod_{t=1}^T\frac{\tr(F_{s_t}^{u_t}\psi^{\otimes 2})}{\tr(F_{s_t}^{u_t}\rho_m^{\otimes 2})}=\frac{\tr\Bigl(\bigotimes_{t=1}^TF_{s_t}^{u_t}\tr_{>n}^{\otimes 2T}(|H\rangle\langle H|^{\otimes 2T})\Bigr)}{\tr\Bigl(\bigotimes_{t=1}^TF_{s_t}^{u_t}\rho_m^{\otimes 2T}\Bigr)}=\frac{\tr\Bigl(\bigotimes_{t=1}^T\tr_{>n}^{\otimes 2\dagger}(F_{s_t}^{u_t})|H\rangle\langle H|^{\otimes 2T}\Bigr)}{\tr\Bigl(\bigotimes_{t=1}^TF_{s_t}^{u_t}\rho_m^{\otimes 2T}\Bigr)},\]
where $\tr_{>n}^{\otimes x}$ means tracing out the last $\log(1/\epsilon)$ qubits for all $x$ copies and $\tr_{>n}^{\otimes x\dagger}$ is its adjoint.
Then the likelihood ratio is
\begin{align*}
L(\ell)&=\E_{\psi\sim\cD_{n,\epsilon}}[L_\psi(\ell)]\\
&=\E_{|H\rangle\sim\mu}\frac{\tr\Bigl(\bigotimes_{t=1}^T\tr_{>n}^{\otimes 2\dagger}(F_{s_t}^{u_t})|H\rangle\langle H|^{\otimes 2T}\Bigr)}{\tr\Bigl(\bigotimes_{t=1}^TF_{s_t}^{u_t}\rho_m^{\otimes 2T}\Bigr)}\\
&=\frac{d^{2T}}{D(D+1)\cdots(D+2T-1)}\frac{\tr\Bigl(\bigotimes_{t=1}^T\tr_{>n}^{\otimes2\dagger}(F_{s_t}^{u_t})\sum_{\pi\in S_{2T}}\pi^D\Bigr)}{\tr\Bigl(\bigotimes_{t=1}^TF_{s_t}^{u_t}\Bigr)}\\
&\geq\frac{d^{2T}}{D(D+1)\cdots(D+2T-1)}\prod_{t=1}^T\frac{\tr(\tr_{>n}^{\otimes2\dagger}(F_{s_t}^{u_t})\sum_{\pi\in S_2}\pi^D)}{\tr(F_{s_t}^{u_t})}\\
&=\frac{d^{2T}}{D(D+1)\cdots(D+2T-1)}\prod_{t=1}^T\frac{\tr(F_{s_t}^{u_t}\tr_{>n}^{\otimes2}(\sum_{\pi\in S_2}\pi^D))}{\tr(F_{s_t}^{u_t})}\\
&=\frac{D^{2T}}{D(D+1)\cdots(D+2T-1)}\prod_{t=1}^T\frac{\tr(F_{s_t}^{u_t}(I+\frac{d}{D}\SWAP))}{\tr(F_{s_t}^{u_t})}\\
&\geq\Bigl(1-\frac{2T}{D}\Bigr)^{2T}\Bigl(1+\frac{1}{D}\Bigr)^T\prod_{t=1}^T\frac{\tr(F_{s_t}^{u_t}\rho_2)}{\tr(F_{s_t}^{u_t}\rho_m^{\otimes2})}\\
&\geq\Bigl(1-\frac{4T^2}{D}\Bigr)\prod_{t=1}^T\frac{\tr(F_{s_t}^{u_t}\rho_2)}{\tr(F_{s_t}^{u_t}\rho_m^{\otimes2})}.
\end{align*}
In the third line we used \Cref{lem:Haar_avg}.
The fourth line uses \Cref{lem:perm_ineq} recursively.
We defined the $2n$-qubits state $\rho_2=\frac{I+d/D\SWAP}{d^2(1+1/D)}$ in the seventh line.
Note that $\prod_{t=1}^T\frac{\tr(F_{s_t}^{u_t}\rho_2)}{\tr(F_{s_t}^{u_t}\rho_m^{\otimes2})}$ is the likelihood ratio of distinguishing $\rho_m^{\otimes2}$ and $\rho_2$ when using $(1,\cM_{2,n}^k)$ protocols.
We can bound it using \Cref{lem: toolbox for lower bounds}(c).
To apply \Cref{lem: toolbox for lower bounds}(c), we need to bound the likelihood ratio at each step.
\begin{align*}
\max_{\{F_s\}_s\in\cM_{2,n}^k}\E_{s\sim\tr(F_s\rho_m^{\otimes2})}\Bigl[\Bigl(\frac{\tr(F_s\rho_2)}{\tr(F_s\rho_m^{\otimes2})}-1\Bigr)^2\Bigr]&=\max_{\{F_s\}_s\in\cM_{2,n}^k}\sum_s\frac{\tr(F_s)}{d^2}\Bigl(\frac{d}{D+1}\frac{\tr(F_s\SWAP)}{\tr(F_s)}-\frac{1}{D+1}\Bigr)^2\\
&\leq\frac{2}{(D+1)^2}+2\max_{\{F_s\}_s\in\cM_{2,n}^k}\frac{\tr(F_s\SWAP)^2}{(D+1)^2\tr(F_s)}\\
&\leq\frac{2}{(D+1)^2}+\frac{2^{k+n+1}}{(D+1)^2}\leq4\epsilon^22^{k-n}.
\end{align*}
In the second line, we used the inequality $(x-y)^2\leq2x^2+2y^2$.
Thus by \Cref{lem: toolbox for lower bounds}(c) we have
\begin{equation*}
\Pr_{\ell\sim p^{\rho_m^{\otimes2}}(\ell)}\Bigl[\prod_{t=1}^T\frac{\tr(F_{s_t}^{u_t}\rho_2)}{\tr(F_{s_t}^{u_t}\rho_m^{\otimes2})}>0.9\Bigr]\geq 0.9-cT(4\epsilon^22^{k-n}),
\end{equation*}
for a constant $c$.
This means that
\begin{equation*}
\Pr_{\ell\sim p^{\rho_m^{\otimes2}}(\ell)}\Bigl[L(\ell)>0.9\Bigl(1-\frac{4T^2}{D}\Bigr)\Bigr]\geq 0.9-cT(4\epsilon^22^{k-n}).
\end{equation*}
Assuming $T\leq\min\{\frac{2^{n/2}}{20\sqrt{\epsilon}},\frac{2^{n-k}}{400c\epsilon^2}\}$, we have $0.9(1-4T^2/D)\geq0.9\times0.99\geq0.89$ and $cT(4\epsilon^22^{k-n})\leq0.01$.
Thus in this case
\[\Pr_{\ell\sim p^{\rho_m^{\otimes2}}(\ell)}[L(\ell)>0.89]\geq0.89.\]
By \Cref{lem: toolbox for lower bounds}(a) and (b), the success probability is at most $0.11+0.11=0.22<\frac{7}{12}$.
We conclude that to solve \Cref{prob: distinguishing} with probability $\frac{7}{12}$, $T\geq\min\{\frac{2^{n/2}}{20\sqrt{\epsilon}},\frac{2^{n-k}}{400c\epsilon^2}\}=\Omega(2^{n/2}/\sqrt{\epsilon},2^{n-k}/\epsilon^2)$.
\end{proof}

\subsection{Improved lower bound restricted to protocols with fixed measurements}\label{sec:improved_lower}

The analysis above gives a $\Omega(\med\{1/\epsilon^2, 2^{n/2}/\sqrt{\epsilon}, 2^{n-k}/\epsilon^2\})$ lower bound. 
There is a $\sqrt{\epsilon}$ gap between this lower bound and the $O(\med\{1/\epsilon^2, 2^{n/2}/\epsilon, 2^{n-k}/\epsilon^2\})$ upper bound that we will show in the next section. We believe that the gap is due to some technique inadequacy in the lower bound proof. Here, we provide a piece of evidence by improving the lower bound in a restricted setting: estimating purity with a randomly fixed PVM.

\begin{definition}[Estimating purity with a randomly fixed single-copy PVM]\label{def: estimate with a randomly fixed PVM}
    We say an algorithm $\mathcal{A}$ estimates the purity within additive error $\epsilon$ with a randomly fixed single-copy PVM, if it proceeds as the following: Given access to an unknown state $\rho$, $\mathcal{A}$ samples a unitary $U$ from a well-designed distribution $\mathcal{U}$ and performs the single-copy PVM $M_U\coloneqq \{U\ketbra{x}U^\dagger\}_{x=0}^{d-1}$ on $T$ copies of $\rho$. It then classically processes all outcomes and outputs an estimator $\hat{E}$ such that with probability at least $2/3$, $\abs{\hat{E}-\tr(\rho^2)}\leq \epsilon$. $T$ is called the sample complexity of $\mathcal{A}$.
\end{definition}

%Specifically, we focus on the single-copy scenario (i.e., $k=0$). We allow the learner to sample a unitary $U$ from a distribution $\mathcal{U}$ designed beforehand. Once the $U$ is sampled, the learner can only perform the PVM $M_U\coloneqq \{U\ketbra{x}U^\dagger\}_{x=0}^{d-1}$ on $\rho$ all the time. We thus call the setting ``learning with a randomly fixed PVM''. Although the setting looks too restrictive, it captures several known algorithms on distributed inner product estimation and purity estimation~\cite{elben2020cross,anshu2022distributed}. 

%Although the setting looks too restrictive, it captures several known algorithms on distributed inner product estimation and purity estimation~\cite{elben2020cross,anshu2022distributed}. (move to intro)

In the single-copy scenario (i.e., $k=0$), one can verify that the aforementioned lower and upper bound simplify to $\Omega(\max\{1/\epsilon^2, 2^{n/2}/\sqrt{\epsilon}\})$ and $O(\max\{1/\epsilon^2, 2^{n/2}/\epsilon\})$, respectively. Now we prove that in the setting of estimating with a randomly fixed PVM, the lower bound can be improved to $\Omega(\max\{1/\epsilon^2, 2^{n/2}/\epsilon\})$, matching the upper bound. 

\begin{theorem}\label{thm:improved_lower}
    Any algorithm $\mathcal{A}$ that estimates the purity within additive error $\epsilon$ with a randomly fixed single-copy PVM (as in \Cref{def: estimate with a randomly fixed PVM}) has sample complexity at least $\Omega(\max\{1/\epsilon^2, 2^{n/2}/\epsilon\})$.
\end{theorem}
\begin{proof}
    The $\Omega(1/\epsilon^2)$ lower bound comes directly from \Cref{lem:entangled_lower}. Here we only need to prove the $\Omega(2^{n/2}/\epsilon)$ lower bound when $1/\epsilon\leq 2^{n/2}$.
    By \Cref{lem:purity_to_dist}, it suffices to prove a lower bound for \Cref{prob: distinguishing}.
    %Denote $d=2^n$ and $D=d/\epsilon$.
    Fix a unitary $U$ sampled from $\mathcal{U}$. We index a leaf $\ell\in \mathrm{Leaf}(\cT)$ by outcomes $(x_1, x_2,\cdots, x_T)\in \{0, 1, \cdots, d-1\}^{T}$ and define $\ket{\ell}\coloneqq \ket{x_1,\cdots, x_T}$ We calculate the probabilities of $\ell$ in both cases.
    \begin{equation}
        p(\ell)\coloneqq p^{\rho_m}(\ell)=\braket{\ell|(U^\dagger)^{\otimes T}\rho_m^{\otimes T}U^{\otimes T}|\ell}=d^{-T}. \label{eq: same PVM p}
    \end{equation}
    \begin{align}
        q(\ell)\coloneqq \E_{\psi\sim \cD_{n, \epsilon}}p^\psi(\ell)&=\braket{\ell|(U^\dagger)^{\otimes T}\E_{\psi}\psi^{\otimes T}U^{\otimes T}|\ell}=\braket{\ell|\E_{\psi}\psi^{\otimes T}|\ell}. \label{eq: same PVM q}
    \end{align}
    Here we use the fact that the distribution of $U^\dagger \psi U$ is the same as the distribution of $\psi$ (i.e., Haar measure is invariant under unitary multiplication). Therefore, the distribution $p$ and $q$ are indeed independent of $U$. The goal is to upper bound the success probability by \Cref{lem: toolbox for lower bounds}(b) when the depth of the learning tree $T\ll \sqrt{d}/\epsilon$. 

    We now explicitly calculate $q(\ell)$ using \Cref{lem:Haar_avg}.
    Here we explicitly specify the local dimension of a permutation operator. For a permutation $\pi\in S_T$, let $\pi^D$ be the permutation operator on $T$ copies, $\pi^{d}$ be the permutation operator on the first $n$ qubits of $T$ copies, and $\pi^{1/\epsilon}$ be the permutation operator on the last $\log(1/\epsilon)$ qubits of $T$ copies. Then we have $\pi^D=\pi^{d}\otimes \pi^{1/\epsilon}$ %(in this equality, $\pi^d$ acts on the first $n$ qubits of all copies, and $\pi^{1/\epsilon}$ acts on the last $\log(1/\epsilon)$ qubits of all copies). 
    Let $c(\pi)$ be the number of cycles in $\pi$. By \Cref{lem:Haar_avg},
    \begin{align}
        \begin{split}\label{eq: same PVM haar}
        \E_{\psi\sim \cD_{n, \epsilon}}\psi^{\otimes T}&=\frac{\tr_{>n}^{\otimes T}(\sum_{\pi\in S_T}\pi^D)}{D(D+1)\cdots (D+T-1)}\\
        &=\frac{\sum_{\pi \in S_T}\pi^d\tr(\pi^{1/\epsilon})}{D(D+1)\cdots (D+T-1)}\\
        &=\frac{\sum_{\pi \in S_T}\pi \epsilon^{-c(\pi)}}{D(D+1)\cdots (D+T-1)}\\
        &=d^{-T}\frac{\sum_{\pi \in S_T}\pi \epsilon^{T-c(\pi)}}{(1+\epsilon/d)\cdots (1+(T-1)\epsilon/d)}.
        \end{split}
    \end{align}
    Here $\tr_{>n}^{\otimes T}$ means tracing out the last $\log(1/\epsilon)$ qubits for all copies. We use \Cref{lem:Haar_avg} in the first line, $\pi^D=\pi^d\otimes \pi^{1/\epsilon}$ in the second line, and $\tr(\pi^{1/\epsilon})=\epsilon^{-c(\pi)}$ in the third line. We omit the superscript $d$ and write $\pi^d$ as $\pi$ from now on. Notice that for any term $\braket{\ell|\pi|\ell}$ to be nonzero, $\pi$ must connect copies with the same outcomes, that is, $x_i=x_{\pi(i)}$ for any $i$. Define $B^{\ell}_y\coloneqq \{i\in \{1, 2, \cdots, T\}:x_i=y\}$ for $y\in \{0,1 , \cdots, d-1\}$, $b^{\ell}_y\coloneqq |B^{\ell}_y|$, and let $S_{B_y^\ell}$ be the set of permutations over copies in $B_y^\ell$. $\braket{\ell|\sum_{\pi\in S_T}\pi \epsilon^{T-c(\pi)}|\ell}$ has a block structure as follows
    \begin{equation}
        \braket{\ell|\sum_{\pi\in S_T}\pi \epsilon^{T-c(\pi)}|\ell}=\prod_{y=0}^{d-1}\braket{y^{\otimes b^\ell_y}|\sum_{\pi\in B_{y}^\ell}\pi \epsilon^{b_y^\ell-c(\pi)}|y^{\otimes b^\ell_y}}=\prod_{y=0}^{d-1}\sum_{\pi\in B_y^\ell}\epsilon^{b_y^\ell-c(\pi)}. \label{eq: same PVM block}
    \end{equation}
    Recall that the coefficient of $x^k$ in $x(x+1)\cdots (x+t-1)$ is the unsigned Stirling number of the first kind, which is the number of permutations of $t$ elements with $k$ cycles (see, e.g. Ref.~\cite{grahamConcreteMathematicsFoundation2017}). So $x(x+1)\cdots (x+t-1)=\sum_{\pi\in S_t}x^{c(\pi)}$. Setting $x=1/\epsilon$, we obtain that $\sum_{\pi\in B_y^\ell}\epsilon^{b_y^\ell-c(\pi)}=(1+\epsilon)(1+2\epsilon)\cdots (1+(b^\ell_y-1)\epsilon)$. Combining with \eqref{eq: same PVM p}, \eqref{eq: same PVM q}, \eqref{eq: same PVM haar}, and \eqref{eq: same PVM block}, we finally obtain the likelihood ratio
    \begin{equation*}
        L(\ell)\coloneqq \frac{q(\ell)}{p(\ell)}=\frac{\prod_{y=0}^{d-1}(1+\epsilon)\cdots (1+(b_y^\ell -1)\epsilon)}{(1+\epsilon/d)\cdots (1+(T-1)\epsilon/d)}.
    \end{equation*}
    Since $x-\frac{x^2}{2}\leq \ln(1+x)\leq x$ for $x\geq0$,
    \begin{align}
        \ln(L(\ell))&= \sum_{y=0}^{d-1}\sum_{i=1}^{b_y^\ell-1}\ln(1+i\epsilon)-\sum_{i=1}^{T-1}\ln(1+i\epsilon/d)\nonumber\\
        &\ge \sum_{y=0}^{d-1}\sum_{i=1}^{b_y^\ell -1}(i\epsilon-\frac{(i\epsilon)^2}{2})-\sum_{i=1}^{T-1}\frac{i\epsilon}{d}\nonumber\\
        &=\epsilon\sum_{y=0}^{d-1}\binom{b_y^\ell}{2}-\frac{\epsilon}{d}\binom{T}{2}-\epsilon^2\sum_{y=0}^{d-1}\frac{(b_y^\ell-1)b_y^\ell(2b_y^\ell-1)}{12}\nonumber\\
        &=\left(\epsilon-\frac{1}{2}\epsilon^2\right)\sum_{y=0}^{d-1}\binom{b_y^\ell}{2}-\frac{\epsilon}{d}\binom{T}{2}-\epsilon^2\sum_{y=0}^{d-1}\binom{b_y^\ell}{3}\nonumber\\
        &=\left(\epsilon-\frac{1}{2}\epsilon^2\right)\sum_{i<j}1[x_i=x_j]-\frac{\epsilon}{d}\binom{T}{2}-\epsilon^2\sum_{i<j<k}1[x_i=x_j=x_k] \label{eq: same PVM count}.
    \end{align}
    We show in \Cref{lem: same PVM counting} that, assuming $T<\sqrt{d}/(100\epsilon)$ and $1/\epsilon<\sqrt{d}$, the right-hand side of \eqref{eq: same PVM count} is larger than $-0.13$ with probability at least $0.97$. In other words, $\Pr_{\ell\sim p(\ell)}[L(\ell)\leq e^{-0.115}]\leq 0.03$. By \Cref{lem: toolbox for lower bounds}(b), the success probability is at most $0.03+(1-e^{-0.115})<2/3$. Therefore, to achieve a $2/3$ success probability, $T=\Omega(\max\{1/\epsilon^2, \sqrt{d}/\epsilon\})$.
\end{proof}

\begin{lemma}\label{lem: same PVM counting}
    Assume $T<\sqrt{d}/(100\epsilon)$ and $1/\epsilon<\sqrt{d}$. Suppose $x_1,x_2,\cdots, x_T$ are uniformly drawn from $\{0, 1, \cdots, d-1\}$. With probability at least $0.97$, the right-hand side of \eqref{eq: same PVM count} is at least $-0.13$.
\end{lemma}
\begin{proof}
    Define two random variables $X\coloneqq \sum_{i<j}1[x_i=x_j], Y\coloneqq \sum_{i<j<k}1[x_i=x_j=x_k]$. It is easy to see $\E[X]=\frac{1}{d}\binom{T}{2}<\frac{1}{10000\epsilon^2}$ and $\E[Y]=\frac{1}{d^2}\binom{T}{3}<\frac{1}{10000\epsilon^3\sqrt{d}}<\frac{1}{10000\epsilon^2}$. Furthermore, for $i_1<j_1, i_2<j_2$
    \begin{equation*}
        \E(1[x_{i_1}=x_{j_1}]1[x_{i_2}=x_{j_2}]])-\E(1[x_{i_1}=x_{j_1}])\E(1[x_{i_2}=x_{j_2}])=\begin{cases}
            \frac{1}{d}-\frac{1}{d^2},~&i_1=i_2,j_1=j_2\\
            0,~&\text{otherwise}.
        \end{cases}.
    \end{equation*}
    So $\Var[X]=\E[X^2]-\E[X]^2=(\frac{1}{d}-\frac{1}{d^2})\binom{T}{2}<\frac{1}{d}\binom{T}{2}<\frac{1}{10000\epsilon^2}$.
    
    By Chebyshev's inequality, with probability at least 0.99, $\abs{X-\E[X]}\leq 10\sqrt{\Var[X]}<\frac{1}{10\epsilon}$. By Markov's inequality, with probability at least 0.99, $X\leq 100\E[X]<\frac{1}{100\epsilon^2}$. With probability at least 0.99, $Y\leq 100\E[Y]<\frac{1}{100\epsilon^2}$. Therefore, with probability at least $0.97$, all events happen and the right-hand side of \eqref{eq: same PVM count} is 
    \begin{equation*}
        \epsilon(X-\E[X])-\frac{1}{2}\epsilon^2 X-\epsilon^2Y>-0.115\,.\qedhere
    \end{equation*}
\end{proof}

\section{Sample complexity of inner product estimation}

\subsection{Inner product estimation with bounded quantum communication}\label{sec:partial_swap}
In this subsection, we provide an algorithm for inner product estimation with $k$-qubit one-way quantum communication (\Cref{prob:inner_product_k}), and thus an algorithm for purity estimation with $k$-qubit quantum memory (\Cref{prob:purity_k}) by \Cref{lem:connection}. 

When $k=0$ in \Cref{prob:inner_product_k}, the task reduces to distributed inner product estimation with unentangled measurements where Alice and Bob estimate $\tr(\rho\sigma)$ using single-copy local measurements and classical communication (i.e., LOCC). In this case, Ref.~\cite{anshu2022distributed} propose a collision-based algorithm using single-copy local measurements and classical communications between Alice and Bob as in \Cref{algo:distributed_ip}.

\begin{algorithm}[ht]
\caption{Distributed inner product estimation using unentangled local measurement~\cite{anshu2022distributed}.}
\label{algo:distributed_ip}
\begin{algorithmic}[1]
\REQUIRE Accuracy demand $\epsilon$, $N$ copies of unknown $\rho$ and $\sigma$.
\ENSURE An estimation of $\tr(\rho\sigma)$ within additive error $\epsilon$.
\STATE{Divide the $N$ copies of $\rho$ and $\sigma$ into $N_b$ groups, each containing $m$ copies of $\rho$ and $\sigma$.}
\FOR{$i=1,...,N_b$}
\STATE{Alice and Bob sample a $(d=2^n)$-dimensional Haar random unitary $U_i$, and load a new batch of $m$ copies of $\rho$ and $\sigma$, respectively.}
\STATE{Alice measures $m$ copies of $\rho$ in the basis $\{U_i^\dagger\ket{b}\bra{b}U\}_{b=0}^{d-1}$ to obtain outputs $x_1,...,x_m$. Alice then sends these classical outputs to Bob.}
\STATE{Bob measures $m$ copies of $\sigma$ in the basis $\{U_i^\dagger\ket{b}\bra{b}U\}_{b=0}^{d-1}$ to obtain outputs $y_1,...,y_m$.}
\STATE{Bob computes the collision estimator
\begin{align*}
\tilde{g}_i=\frac{1}{m^2}\sum_{j,k=1}^m\mathbbm{1}[x_j=y_k].
\end{align*}
}
\STATE{Let $w_i=(2^n+1)\tilde{g_i}-1$.}
\ENDFOR
\STATE{Bob outputs $w=\frac{1}{N_b}\sum_{i=1}^{N_b}w_i$.}
\end{algorithmic}
\end{algorithm}

Here, we write the protocol in a one-way communication version. The following lemma proves the performance of \Cref{algo:distributed_ip}: 
\begin{lemma}[Performance of \Cref{algo:distributed_ip},~\cite{anshu2022distributed}]\label{lem:algo_distributed_ip}
There exists a choice of $N_b$ and $m$ such that 
\begin{align*}
N=N_bm=O\left(\max\left\{\frac{1}{\epsilon^2},\frac{2^{n/2}}{\epsilon}\right\}\right),
\end{align*}
and the estimator $w$ output from \Cref{algo:distributed_ip} is an approximation of $\tr(\rho\sigma)$ within additive error $\epsilon$ with a high probability.
\end{lemma}

In \Cref{prob:inner_product_k}, Alice and Bob are allowed to have $k$-qubit one-way quantum communication for each copy pair $\rho$ and $\sigma$. To this end, we propose the following partial swap test algorithm making use of one-way quantum communication in \Cref{algo:partial_swap}.

\begin{algorithm}[ht]
\caption{Partial swap test using $k$-qubit one-way communication unentangled local measurement.}
\label{algo:partial_swap}
\begin{algorithmic}[1]
\REQUIRE Accuracy demand $\epsilon$, $N+N_k$ copies of unknown $\rho$ and $\sigma$ where $N_k=O(1/\epsilon^2)$.
\ENSURE An estimation of $\tr(\rho\sigma)$ within additive error $\epsilon$.
\STATE{Alice and Bob load $N_k$ copies of $\rho$ and $\sigma$, respectively.}\label{line:partial_trace_start}
\STATE{Alice sends the first $k$ qubits of each copy to Bob.}
\STATE{Bob applies a partial swap test, i.e. POVM $\left\{\frac{I_{2k}+\SWAP_k}{2},\frac{I_{2k}-\SWAP_k}{2}\right\}$, on the first $k$ qubits of $\rho$ received and $\sigma$ for each copy. He can then estimate $\tilde{f}_k$ within $\epsilon/2$ of the inner product $f_k=\tr(\tr_{>k}(\rho)\tr_{>k}(\sigma))$.}\label{line:partial_trace_end}
\STATE{Divided the $N$ copies of $\rho$ and $\sigma$ into $N_b$ groups, each containing $m$ copies of $\rho$ and $\sigma$.}
\FOR{$i=1,...,N_b$}
\STATE{Alice and Bob sample a $2^{n-k}$-dimensional Haar random unitary $U_i$, and load a new batch of $m$ copies of $\rho$ and $\sigma$, respectively.}
\STATE{Alice sends the first $k$ qubits of each copy to Bob.}
\STATE{Bob applies a partial swap test, i.e. POVM $\left\{\frac{I+\SWAP_k}{2},\frac{I-\SWAP_k}{2}\right\}$, on the first $k$ qubits of $\rho$ received and $\sigma$ for each copy, and obtains outcome $z_1,...,z_m$.}
\STATE{Alice measures $m$ copies of the remaining $n-k$ qubits of $\rho$ each in the basis $\{U_i^\dagger\ket{b}\bra{b}U\}_{b=0}^{2^{n-k}-1}$ to obtain outputs $x_1,...,x_m$. Alice then sends these classical outputs to Bob.}
\STATE{Bob measures $m$ copies of the remaining $n-k$ qubits of $\sigma$ each in the basis $\{U_i^\dagger\ket{b}\bra{b}U\}_{b=1}^{2^{n-k}-0}$ to obtain outputs $y_1,...,y_m$.}
\STATE{Bob computes the value
\begin{align*}
\tilde{g}_i=\frac{1}{m}\sum_{i=1}^m\left(\mathbbm{1}[z_i=1,x_i=y_i]-\mathbbm{1}[z_i=-1,x_i=y_i]\right)
\end{align*}
}
\STATE{Let $w_i=(2^{n-k}+1)\tilde{g_i}-\tilde{f}_k$}
\ENDFOR
\STATE{Bob outputs $w=\frac{1}{N_b}\sum_{i=1}^{N_b}w_i$}
\end{algorithmic}
\end{algorithm}

Now, we analyze the performance of \Cref{algo:partial_swap}. From Line~\ref{line:partial_trace_start} to Line\ref{line:partial_trace_end}, Alice and Bob are to perform the folklore swap test on the first $k$ qubit of $\rho$ and $\sigma$. By choosing $N_k=O(1/\epsilon^2)$, one can guarantee that Alice and Bob can estimate $\tilde{f}_k$ within $\epsilon/2$ of the inner product $f_k=\tr(\tr_{>k}(\rho)\tr_{>k}(\sigma))$ with high probability.

We then consider the remaining part of the algorithm. In each iteration $i=1,...,N_b$, we denote $X=\{x_1,...,x_m\}$, $Y=\{y_1,...,y_m\}$, $Z=\{z_1,...,z_m\}$, and $S=\{X,Y,Z\}$. We can thus regard $\tilde{g}_i$ as a random function $g(U,S)$. Note that fix $U$, we have
\begin{align*}
\Pr[z_i=1,x_i=y_i]&=\sum_{b=0}^{d-1}\tr\left(\frac{I_{2k}+\SWAP_k}{2}\otimes \left(U^\dagger\ket{b}\bra{b} U\right)^{\otimes 2}\cdot\rho\otimes\sigma\right),\\
\Pr[z_i=-1,x_i=y_i]&=\sum_{b=0}^{d-1}\tr\left(\frac{I_{2k}-\SWAP_k}{2}\otimes \left(U^\dagger\ket{b}\bra{b} U\right)^{\otimes 2}\cdot\rho\otimes\sigma\right).
\end{align*}
The expectation value of $g(U,S)$ is thus
\begin{align*}
\underset{U,S}{\mathbb{E}}[g(U,S)]&=\underset{U}{\mathbb{E}}\left[\sum_{b=0}^{d-1}\tr\left(\SWAP_k\otimes \left(U^\dagger\ket{b}\bra{b} U\right)^{\otimes 2}\cdot\rho\otimes\sigma\right)\right]\\
&=\frac{1}{2^{n-k}+1}\tr\left(\SWAP_k\otimes\left(I_{n-k}\otimes I_{n-k}+\SWAP_{n-k}\right)\cdot \rho\otimes\sigma\right)\\
&=\frac{f+f_k}{2^{n-k}+1},
\end{align*}
where $f_k=\tr(\tr_{>k}(\rho)\tr_{>k}(\sigma))$ and $f=\tr(\rho\sigma)$. This implies the following:
\begin{lemma}\label{lem:average_partial_trace}
The value $w$ output from Bob in \Cref{algo:partial_swap} is an unbiased estimation for $\tr(\rho\sigma)$.
\end{lemma}

Note that since Alice and Bob can estimate $\tilde{f}_k$ within $\epsilon/2$ of the inner product $f_k=\tr(\tr_{>k}(\rho)\tr_{>k}(\sigma))$ with high probability, we only need to compute the parameters $N$, $N_b$, and $m$ such that the remaining part of $w$ deviates from $\tr(\rho\sigma)+f_k$ for at most $\epsilon/2$. To achieve this, we consider the variance of $w$. Note that 
\begin{align}\label{eq:variance_w}
\Var(w)=\frac{1}{N_b^2}\sum_{i=1}^{N_b}\Var(w_i)=\frac{(2^{n-k}+1)^2}{N_b}\Var(g(U,S)),
\end{align}
it is sufficient to compute the variance of $g(U,S)$. The variance of $\Var(g(U,S))$ can be divided into the variance of the random unitary $U$ and the variance of the intrinsic randomness of quantum measurement. This indicates that
\begin{align}\label{eq:variance_g_total}
\Var(g(U,S))=\Var_U(\mathbb{E}[g(U,S)|U])+\mathbb{E}_U[\Var(g(U,S)|U)].
\end{align}

We now compute the two terms. We have the following lemma for the first term:
\begin{lemma}\label{lem:variance_first}
The first term in \eqref{eq:variance_g_total} is bounded by:
\begin{align*}
\Var_U(\mathbb{E}[g(U,S)|U])=O\left(\frac{1}{2^{3(n-k)}}\right).
\end{align*}
\end{lemma}

\begin{proof}
The first term in \eqref{eq:variance_g_total} can be computed as
\begin{align}\label{eq:variance_g_1}
\begin{split}
&\quad\Var_U(\mathbb{E}[g(U,S)|U])\\
&=\underset{U}{\mathbb{E}} [g(U)^2]-\underset{U}{\mathbb{E}}[g(U)]^2\\
&=\sum_{b_1,b_2=0}^{2^{n-k}-1}\underset{U}{\mathbb{E}}\left[\tr\left((\SWAP_k)^{\otimes 2}\otimes \left(U^\dagger\ket{b_1}\bra{b_1} U\right)^{\otimes 2}\otimes\left(U^\dagger\ket{b_2}\bra{b_2} U\right)^{\otimes 2}\cdot(\rho\otimes\sigma)^{\otimes 2}\right)\right]-\underset{U}{\mathbb{E}}[g(U)]^2.
\end{split}
\end{align}
While the last term above is $\mathbb{E}_U[g(U)]^2=\left(\frac{f+f_k}{2^{n-k}+1}\right)^2$, the first term can be further decomposed into the case when $b_1=b_2$ and $b_1\neq b_2$. For $b_1=b_2$, the total contribution can be computed as
\begin{align}\label{eq:variance_g_1_1}
\begin{split}
&\quad 2^{n-k}\underset{\psi\sim \mu}{\mathbb{E}}\left[\tr\left((\SWAP_k)^{\otimes 2}\otimes(\ket{\psi}\bra{\psi})^{\otimes 4}\cdot(\rho\otimes\sigma)^{\otimes 2}\right)\right]\\
&=\frac{1}{(2^{n-k}+1)(2^{n-k}+2)(2^{n-k}+3)}\sum_{\pi\in S_4}\tr\left((\SWAP_k)^{\otimes 2}\otimes\pi^{2^{n-k}}\cdot(\rho\otimes\sigma)^{\otimes 2}\right)\\
&=O\left(\frac{1}{2^{3(n-k)}}\right),  
\end{split}
\end{align}
where the second line follows from \Cref{lem:Haar_avg} and the last line follows from the fact that the summation is over $24$ elements each of at most $1$. When $b_1\neq b_2$, the contribution can be computed as
\begin{align*}
\quad 2^{n-k}(2^{n-k}-1)\mathbb{E}_{\psi\sim\mu\atop \psi'\sim\mu_{\perp\psi}}\tr\left((\SWAP_k)^{\otimes 2}\otimes \left(\ket{\psi}\bra{\psi}\right)^{\otimes 2}\otimes\left(\ket{\psi'}\bra{\psi'}\right)^{\otimes 2}\cdot(\rho\otimes\sigma)^{\otimes 2}\right).
\end{align*}
Here $\mu_{\perp\psi}$ is the Haar measure over states perpendicular to $|\psi\rangle$.
We have the following lemma to deal with the above equation:
\begin{lemma}[See Lemma 18 of Ref.~\cite{anshu2022distributed}]
\begin{align*}
\underset{\psi'\sim\mu_{\perp\psi}}{\mathbb{E}}\left[\ket{\psi'}\bra{\psi'}^{\otimes 2}\right]=\frac{1}{2^{n-k}(2^{n-k}-1)}\left((I-\ket{\psi}\bra{\psi})^{\otimes 2}+(I-\ket{\psi}\bra{\psi})^{\otimes 2}\SWAP_{n-k}(I-\ket{\psi}\bra{\psi})^{\otimes 2}\right).
\end{align*}
\end{lemma}
By plugging this lemma into the case when $b_1\neq b_2$, it can be written as
\begin{align}\label{eq:variance_g_1_2}
\begin{split}
&\quad 2^{n-k}(2^{n-k}-1)\mathbb{E}_{\psi\sim\mu\atop \psi'\sim\mu_{\perp\psi}}\tr\left((\SWAP_k)^{\otimes 2}\otimes \left(\ket{\psi}\bra{\psi}\right)^{\otimes 2}\otimes\left(\ket{\psi'}\bra{\psi'}\right)^{\otimes 2}\cdot(\rho\otimes\sigma)^{\otimes 2}\right)\\
&=\mathbb{E}_{\psi\sim\mu}\tr\left((\SWAP_k)^{\otimes 2}\otimes \left(\ket{\psi}\bra{\psi}\right)^{\otimes 2}\otimes\left(I\otimes I+\SWAP_{n-k}\right)\cdot(\rho\otimes\sigma)^{\otimes 2}\right)+O\left(2^{-3(n-k)}\right)\\
&=\frac{\tr\left((\SWAP_k)^{\otimes 2}\otimes \left(I\otimes I+\SWAP_{n-k}\right)\otimes\left(I\otimes I+\SWAP_{n-k}\right)\cdot(\rho\otimes\sigma)^{\otimes 2}\right)}{2^{n-k}(2^{n-k}+1)}+O\left(2^{-3(n-k)}\right)\\
&=\frac{(f+f_k)^2}{2^{n-k}(2^{n-k}+1)}+O\left(2^{-3(n-k)}\right).
\end{split}
\end{align}
By inserting \eqref{eq:variance_g_1_1} and \eqref{eq:variance_g_1_2} into \eqref{eq:variance_g_1}, we arrive at 
\begin{align*}
\Var_U(\mathbb{E}[g(U,S)|U])=\frac{(f+f_k)^2}{2^{n-k}(2^{n-k}+1)}+O\left(2^{-3(n-k)}\right)-\left(\frac{f+f_k}{2^{n-k}+1}\right)^2=O\left(2^{-3(n-k)}\right).
\end{align*}
which proves \Cref{lem:variance_first}.
\end{proof}

Regarding the second term, we have:
\begin{lemma}\label{lem:variance_second}
The second term in \eqref{eq:variance_g_total} is bounded by:
\begin{align*}
\mathbb{E}_U[\Var(g(U,S)|U)]=O\left(\frac{1}{m2^{n-k}}\right).
\end{align*}
\end{lemma}

\begin{proof}
We first note that when we fix a $U$,
\begin{align*}
\Var(g)&=\mathbb{E}[g^2]-\mathbb{E}[g]^2\\
&=\frac{1}{m^2}\sum_{i,j=1}^m\left(\mathbbm{1}[z_i=1,x_i=y_i]-\mathbbm{1}[z_i=-1,x_i=y_i]\right)\left(\mathbbm{1}[z_j=1,x_j=y_j]-\mathbbm{1}[z_j=-1,x_j=y_j]\right)-\mathbb{E}[g]^2\\
&=\frac{\mathbb{E}[g']}{m}+\left(\frac{m-1}{m}-1\right)\mathbb{E}[g]^2\\
&\leq\frac{\mathbb{E}[g']}{m},
\end{align*}
Here, $g'=\tilde{g}_i=\frac{1}{m}\sum_{i=1}^m\left(\mathbbm{1}[z_i=1,x_i=y_i]+\mathbbm{1}[z_i=-1,x_i=y_i]\right)$ and the third line follows from considering $i=j$ and $i\neq j$. Averaging over $U$, we have
\begin{equation*}
\mathbb{E}_U[\Var(g(U,S)|U)]=\frac{\mathbb{E}_{U,S}[g']}{m}=O\left(\frac{1}{m2^{n-k}}\right),
\end{equation*}
where second step follows from the fact that
\begin{align*}
\mathbb{E}_{U,S}[g']&=\underset{U}{\mathbb{E}}\left[\sum_{b=0}^{d-1}\tr\left(I_k\otimes I_k\otimes \left(U^\dagger\ket{b}\bra{b} U\right)^{\otimes 2}\cdot\rho\otimes\sigma\right)\right]\\
&=\frac{1}{2^{n-k}+1}\tr\left(I_k\otimes I_k\otimes\left(I_{n-k}\otimes I_{n-k}+\SWAP_{n-k}\right)\cdot \rho\otimes\sigma\right)\\
&\leq\frac{2}{2^{n-k}+1}.
\end{align*}
\end{proof}

Combining \Cref{lem:average_partial_trace}, \Cref{lem:variance_first}, and \Cref{lem:variance_second}, we have the following lemma regarding the performance of \Cref{algo:partial_swap}:
\begin{lemma}[Performance of \Cref{algo:partial_swap}]\label{lem:algo_partial_swap}
There exists a choice of $N_b$ and $m$ such that 
\begin{align*}
N=N_bm=O\left(\frac{2^{n-k}}{\epsilon^2}\right),
\end{align*}
and the estimator $w$ output from \Cref{algo:partial_swap} is an approximation of $\tr(\rho)$ within additive error $\epsilon$ with a high probability.
\end{lemma}

\begin{proof}
We first combine \Cref{lem:variance_first} and \Cref{lem:variance_second}, we can compute the variance in \eqref{eq:variance_g_total} as
\begin{align*}
\Var(g(U,S))=\Var_U(\mathbb{E}[g(U,S)|U])+\mathbb{E}_U[\Var(g(U,S)|U)]=O\left(\frac{1}{m2^{n-k}}+\frac{1}{2^{3(n-k)}}\right).
\end{align*}
Plugging the above equation in \eqref{eq:variance_w}, we have
\begin{align*}
\Var(w)=\frac{(2^{n-k}+1)^2}{N_b}\Var(g(U,S))=\frac{1}{N_b}O\left(\frac{2^{n-k}}{m}+\frac{1}{2^{n-k}}\right).
\end{align*}
Note that from \Cref{lem:average_partial_trace}, $w$ output from Bob is an unbiased estimation for $\tr(\rho\sigma)$. It is enough to satisfy
\begin{align*}
\Var(w)\leq\left(\frac{\epsilon}{2}\right)^2.
\end{align*}
Then we have with high probability, $w+\tilde{f}_k$ deviates from $\tr(\rho\sigma)+f_k$ for at most $\epsilon/2$. As $\tilde{f}_k$ deviates from $f_k$ for at most $\epsilon/2$ with high probability, we conclude that $w$ deviates from $\tr(\rho\sigma)$ for at most $\epsilon$ with high probability. Therefore, we require
\begin{align*}
\frac{2^{n-k}}{N_bm}\leq O\left(\epsilon^2\right),\ \frac{1}{N_b2^{n-k}}\leq O\left(\epsilon^2\right)
\end{align*}
Here, we set $m$ to $1$, and $N_b=O(2^{n-k}/\epsilon^2)$ satisfies the requirements. Thus, we have
\begin{equation*}
%N=N_bm+N_k=O\left(\max\left\{\frac{1}{\epsilon^2},\frac{2^{n-k}}{\epsilon^2}\right\}\right)\,.\qedhere
N = N_bm = O\left(\frac{2^{n-k}}{\epsilon^2}\right).\qedhere
\end{equation*}
\end{proof}

Therefore, the sample complexity of \Cref{algo:partial_swap} is $N+N_k=O(2^{n-k}/\epsilon^2)$. Combining \Cref{algo:distributed_ip} and \Cref{algo:partial_swap}, we reach the following theorem:
\begin{theorem}\label{thm:partial_swap}
Given unknown quantum states $\rho$ and $\sigma$, error parameter $\epsilon$, and the capacity of one-way communication $k$ in \Cref{prob:inner_product_k}, the combination of \Cref{algo:distributed_ip} and \Cref{algo:partial_swap} yields an algorithm that solves \Cref{prob:inner_product_k} with a high probability using
\begin{align*}
O\left(\min\left\{\frac{2^{n-k}}{\epsilon^2},\max\left\{\frac{1}{\epsilon^2},\frac{2^{n/2}}{\epsilon}\right\}\right\}\right)=O\left(\med\left\{\frac{1}{\epsilon^2},\frac{2^{n/2}}{\epsilon},\frac{2^{n-k}}{\epsilon^2}\right\}\right)
\end{align*}
copies of $\rho$ and $\sigma$.
\end{theorem}

\subsection{Lower bound for mixed state decisional inner product estimation}\label{sec:mix_dipe}

In this subsection, we provide two lower bounds for mixed state versions of the DIPE introduced in Ref.~\cite{anshu2022distributed}.
For both versions, we obtain a sample lower bound of $\sqrt{d/\tr(\rho^2)}$.
For Problem~\ref{problem:DIPEfixed} our bound holds for protocols with one-way communication between Alice and Bob. 
The proof follows directly from concentration of measure and, in particular, does not require Chiribella's theorem. 
For Problem~\ref{thm:DIPEconvex} we can show the lower bound for all separable measurements (see \Cref{def:separable_measurements}) by adapting the proof technique in Ref.~\cite{anshu2022distributed}.

\subsubsection{Lower bounds from concentration of measure}\label{section:lowerboundfixed}
We consider the following mixed variant of the DIPE:
\begin{problem}[Mixed state decisional inner product estimation with fixed spectrum, mixed DIPE I]\label{problem:DIPEfixed}
Fixed an underlying quantum state $\rho$, Alice and Bob are given copies of states promised that one of the following two cases hold:
\begin{itemize}
    \item Alice and Bob both have state $U\rho U^\dagger$ for a Haar randomly chosen unitary $U$.
    \item Alice has state $U\rho U^\dagger$ and Bob has stat $V\rho V^\dagger$ for independent Haar randomly chosen unitary $U$ and $V$.
\end{itemize}
Their goal is to decide which case they are in with high probability.
\end{problem}
\begin{theorem}\label{thm:DIPEfixed}
    Deciding DIPE I for any fixed $\rho$ requires at least $\Omega(\sqrt{d/\tr(\rho^2)})$ copies if Alice and Bob are allowed to perform local measurements and one-way communication protocols. 
\end{theorem}

We need a standard lemma bounding the Lipshitz constant for $k$-copy measurements:

\begin{lemma}
	$f_{U,\rho}:=\tr[M(U\rho U^{\dagger})^{\otimes k}]$ has Lipschitz constant $L\leq 2k\sqrt{\tr[\rho^2]}$ for any operator $M$ with $||M||_{\infty}\leq 1$.
\end{lemma}
\begin{proof}
We calculate
\begin{align*}
	\begin{split}
		|\tr[M(U\rho U^{\dagger})^{\otimes k}]-\tr[M(V\rho V^{\dagger})^{\otimes k}]| &=	|\tr[M((U\rho U^{\dagger})^{\otimes k}-(V\rho V^{\dagger})^{\otimes k})]|\\
		&\leq ||M||_{\infty}||(U\rho U^{\dagger})^{\otimes k}-(V\rho V^{\dagger})^{\otimes k}||_1\\
		&\leq ||(U\rho U^{\dagger})^{\otimes k}-(V\rho V^{\dagger})^{\otimes k}||_1
	\end{split}
\end{align*}
We can now use the following bound:
	Consider matrices $A,B,C,D$ with $||A||_1=||D||_1=1$.
	Then,
	\begin{align*}
		\begin{split}
		||A\otimes B-C\otimes D||_1&\leq ||A\otimes (B-D)+(A-C)\otimes D||_1\\
		&\leq ||A||_1||B-D||_1+||A-C||_1||D||_1\\
		&\leq ||B-D||_1+||A-C||_1.
		\end{split}
	\end{align*}
We apply this bound $k$ times and obtain
\begin{equation*}
	\begin{split}
	|\tr[M(U\rho U^{\dagger})^{\otimes k}]-\tr[M(V\rho V^{\dagger})^{\otimes k}]|&\leq k||U\rho U^{\dagger}-V\rho V^{\dagger}||_1\\
 &=k||U\rho U^{\dagger}- U\rho V^{\dagger}+U\rho V^{\dagger}-V\rho V^{\dagger}||_1\\
 &\leq k||U\rho (U^{\dagger}-V^{\dagger})||_1+k||(U-V)\rho V^{\dagger}||_1\\
 &\leq 2k||\rho||_2 ||U-V||_2.
	\end{split}
\end{equation*}
by Hölder's inequality and Cauchy-Schwarz.
% where we used that $(x-1)^2\geq 0$ implies $\frac14 (1-x^2)\leq \frac12 (1-x)$ for $x=\mathrm{Re}(\langle \psi|\phi\rangle)$.
\end{proof}

We can now prove Theorem~\ref{thm:DIPEfixed}:
\begin{proof}[Proof of Theorem~\ref{thm:DIPEfixed}]
For quantum measurements $M_{A,B}=\sum_{i}M_i\otimes N_i$ performed by Alice and Bob with one-way communications (see Measurements with classical communications part in \Cref{sec:quantum_info} for details), we will show that the assumption
%\zz{We need to define $M_i$ and $N_i$ somewhere.}
\begin{equation*}
	\underset{U,V\sim\mu}{\mathbb{E}}\tr[M_{A,B} (U\rho U^{\dagger})^{\otimes k}\otimes (V\rho V^{\dagger})^{\otimes k}]=\underset{U\sim\mu}{\mathbb{E}} \sum_i\tr[M_i (U\rho U^{\dagger})^{\otimes k}]\underset{V\sim\mu}{\mathbb{E}}\tr[N_i (V\rho V^{\dagger})^{\otimes k}]\geq \frac23
\end{equation*}
leads to a contradiction for $k=o(\sqrt{d/\tr(\rho^2)})$.
Theorem~\ref{thm:levy} implies
\begin{equation*}
	\mathrm{Pr}[|\tr[N_i(V\rho V^{\dagger})^{\otimes k}]-\mathbb{E}_{V}\tr[N_i (V\rho V^{\dagger})^{\otimes k}]|\geq \tau]\leq 4e^{-\frac{2d \tau^2}{36\pi^3\tr[\rho^2]k^2}}.
\end{equation*}
With the notation $B_\tau:=\{|\tr[N_i(V\rho V^{\dagger})^{\otimes k}]-\mathbb{E}_{V}\tr[N_i (V\rho V^{\dagger})^{\otimes k}]|\leq \tau\}$. 
\begin{align*}%\label{eq:oneway_mixed_tech}
	\begin{split}
	&\underset{U\sim\mu}{\mathbb{E}} \tr[M_{A,B}(U\rho U^{\dagger})^{\otimes k}\otimes (U\rho U^{\dagger})^{\otimes k}]\\
	 &= \sum_i\underset{U\sim\mu}{\mathbb{E}}\tr[M_i(U\rho U^{\dagger})^{\otimes k}]\tr[N_i (U\rho U^{\dagger})^{\otimes k}]\\
	 &\geq \sum_i\int_{B_{\tau}}\tr[M_i(U\rho U^{\dagger})^{\otimes k}]\tr[N_i(U\rho U^{\dagger})^{\otimes k}]\mathrm{d}\mu(U)\\
	 & \geq\sum_i \int_{B_{\tau}}\tr[M_i(U\rho U^{\dagger})^{\otimes k}]\left(\underset{V\sim\mu}{\mathbb{E}} \tr[N_i(V\rho V^{\dagger})^{\otimes k}]-\tau\right)\mathrm{d}\mu(U)\\
	 & \geq \underset{U\sim\mu}{\mathbb{E}} \sum_i \tr[M_i(U\rho U^{\dagger})^{\otimes k}]\underset{V\sim\mu}{\mathbb{E}} \tr[N_i(V\rho V^{\dagger})^{\otimes k}] -\tau -\int_{U(d)\setminus B_{\tau}}\sum_i \tr[M_i(U\rho U^{\dagger})^{\otimes k}]\\
	 & \geq \frac23 -\tau -4e^{-\frac{2d \tau^2}{36\pi^3 \tr[\rho^2]k^2}}
	\end{split}
\end{align*}
We can now choose $\tau=1/10$ and find that for $k\leq \sqrt{d}/400\sqrt{\tr[\rho^2]}$ mixed DIPE I cannot be solved with general one-way protocols.
\end{proof}

\subsubsection{Lower bound for all separable measurements}\label{section:convexlowerbound}
We consider a second mixed variant of the DIPE to prove a lower bound even if Alice and Bob are allowed to perform arbitrary measurements.
The following is a restatement of Problem~\ref{problem:DIPEconvexIntro}:

\begin{problem}[Convex mixture of Haar random states, mixed DIPE II]\label{problem:DIPEconvex}
Alice and Bob are given copies of states promised that one of the following two cases hold:
\begin{itemize}
    \item Alice and Bob both have state $\rho=\frac1{r}\sum_{i=1}^r |\psi_i\rangle\langle\psi_i|$, where the states $\psi_i$ are drawn iid from $\mu$.
    \item Alice has state $\rho=\frac1{r}\sum_{i=1}^r |\psi_i\rangle\langle\psi_i|$ and Bob has state $\sigma=\frac1{r}\sum_{i=1}^r |\phi_i\rangle\langle\phi_i|$ where the states $|\phi_i\rangle$ and $|\psi_i\rangle$ are drawn iid from $\mu$.
\end{itemize}
Their goal is to decide which case they are in with high probability.
\end{problem}

We will prove the following Theorem:
\begin{theorem}\label{thm:DIPEconvex}
Any protocol involving arbitrary interactive protocols between Alice and Bob requires $k=\Omega(\sqrt{d/\tr(\rho^2)})$ copies to decide mixed DIPE II with high probability.
\end{theorem}

As in Ref.~\cite{anshu2022distributed}, we can use Chiribella's theorem~\cite{chiribella2010quantum} that characterizes the measure-and-prepare channel via the optimal cloning channel.
The measure-and-prepare channel~\cite{harrow2013church} is defined as
\begin{equation*}
    \mathrm{MP}_{a\to b}(\rho):={a+d-1\choose a}\underset{\psi\sim\mu}{\mathbb{E}}\tr(\rho\cdot\psi^{\otimes a})\psi^{\otimes b}.
\end{equation*}
As we will see, the measure-and-prepare channel will show up naturally when analyzing mixed DIPE II.
On the other hand we have the optimal cloning channel:
\begin{equation*}
\mathrm{Clone}_{a\to b}(\rho):= \frac{{d+a-1\choose a}}{{d+b-1\choose b}} \mathcal{S}_{b}\left(\rho\otimes \mathbbm{1}^{\otimes b-a}\right)\mathcal{S}_b.
\end{equation*}
Chiribella's theorem~\cite{chiribella2010quantum} then establishes the following connection:
\begin{theorem}[Chiribella's theorem]\label{thm:chiribella}
For integers $a,b\geq 0$ and a density matrix $\rho$ it holds that
\begin{equation*}
    \mathrm{MP}_{a\to b}(\rho) = \sum_{s=0}^{\min a,b} \frac{{a\choose s}{d+b-1\choose b-s}}{{d+a+b-1\choose b}}\mathrm{Clone}_{s\to b}(\tr_{a-s}[\rho]).
\end{equation*}
\end{theorem}
Theorem~\ref{thm:chiribella} allows us to prove the following generalization of Lemma 7 in Ref.~\cite{anshu2022distributed}:
\begin{lemma}\label{lemma:bound on post measurement}
For any input state $\rho$ we have 
\begin{equation*}
    \underset{\psi\sim\mu}{\mathbb{E}}\tr[\rho\psi^{\otimes a}]\psi^{\otimes b}\geq e^{-ab/d}\frac{\mathcal{S}_b}{{d+b-1\choose b}}.
\end{equation*}
\end{lemma}
\begin{proof}
    Applying Theorem~\ref{thm:chiribella}, we calculate:
    \begin{align*}
    \begin{split}
       \underset{\psi}{\mathbb{E}}\tr[\rho\psi^{\otimes a}]\psi^{\otimes b}&=\frac{1}{{a+d-1\choose a}}\mathrm{MP}_{a\to b}(\rho)\\
       &=\frac{1}{{a+d-1\choose a}}\sum_{s=0}^{\min a,b} \frac{{a\choose s}{d+b-1\choose b-s}}{{d+a+b-1\choose b}}\mathrm{Clone}_{s\to b}(\tr_{a-s}[\rho])\\
       &\geq\frac{1}{{a+d-1\choose a}} \frac{{a\choose 0}{d+b-1\choose b}}{{d+a+b-1\choose b}}\mathrm{Clone}_{0\to b}(\tr[\rho])\\
       &=\frac{1}{{a+d-1\choose a}}\frac{{d+b-1\choose b}}{{d+a+b-1\choose b}} \frac{\mathcal{S}_b}{{d+b-1\choose b}}\\
       &\geq\frac{1}{{a+d-1\choose a}} e^{-ab/d}\frac{\mathcal{S}_b}{{d+b-1\choose b}},
       \end{split}
    \end{align*}
    where the last line follows from
    \begin{equation*}
    \frac{\prod_{i=0}^{b-1}(d+i)}{\prod_{i=0}^{b-1}(d+a+i)}=\prod_{i=1}^{b-1}\frac{1}{1+\frac{a}{d+i}}\geq e^{-\sum_{i=0}^{b-1}\frac{a}{d+i}}\geq e^{-\frac{ab}{d}}\,.\qedhere
    \end{equation*}
    % \zz{I think now it's $\frac{1}{\binom{a+d-1}{a}}e^{-ab/d}\frac{\mathcal{S}_b}{\binom{d+b-1}{b}}$. Also need to change the RHS of lemma 15 accordingly.}
    % \jh{I could only easily show $e^{-(a^2+ab)/d}$ actually. Will implement the change}
    % \zz{\[\frac{\prod_{i=0}^{b-1}(d+i)}{\prod_{i=0}^{b-1}(d+a+i)}=\prod_{i=1}^{b-1}\frac{1}{1+\frac{a}{d+i}}\geq e^{-\sum_{i=0}^{b-1}\frac{a}{d+i}}\geq e^{-\frac{ab}{d}}\]}
\end{proof}
% \jh{ah, I see. Fair enough... Not sure where the $a^2$ came from this afternoon. Thanks!}

\begin{proof}[Proof of Theorem~\ref{thm:DIPEconvex}]

We will show that a successful separable two-outcome POVM $\{M_{A,B},\mathbbm{1}-M_{A,B}\}$ with $k=o(\sqrt{d/\mathbb{E}\tr(\rho^2)})$ will lead to a contradiction.
More precisely, we assume that $\{M_{A,B},\mathbbm{1}-M_{A,B}\}$ satisfies 
\begin{enumerate}
	\item $\mathbb{E}_{\rho} \tr(M_{A,B}\rho^{\otimes k}\otimes \rho^{\otimes k})\leq \frac13$.
	\item $\mathbb{E}_{\rho,\sigma}\tr(M_{A,B} \rho^{\otimes k}\otimes \sigma^{\otimes k})\geq \frac23$.
\end{enumerate}

\noindent First, notice that
\begin{equation*}
    \underset{\rho}{\mathbb{E}}\tr(\rho^2)= \frac{1}{r^2}\sum_{i,j=1}^r \underset{\psi_1,\ldots,\psi_r\sim\mu}{\mathbb{E}}|\langle \psi_i|\psi_j\rangle|^2=\frac{1}{r^2}(r+r(r-1)d^{-1}),
\end{equation*}
where we used the formula $\mathbb{E}_{\psi} |\langle\psi|\phi\rangle|^2=d^{-1}$ for any fixed state $\phi$.
Therefore, we can assume that $\mathbb{E}_{\rho}\tr(\rho^2)=\Theta(r^{-1})$.

We will consider the (unnormalized) ``post measurement'' state of Bob in case 1:
\begin{equation*}
    \rho_{\mathrm{Bob}}= \underset{\rho}{\mathbb{E}}\tr(A_t \rho^{\otimes k})\rho^{\otimes k}.
\end{equation*}
We expand:
\begin{equation*}
   \underset{\rho}{\mathbb{E}}\tr(A_t\rho^{\otimes k})=\sum_{i_1,...,i_k=1}^r \tr\left[A_t\underset{{\psi_1,\ldots,\psi_r\sim\mu}}{\mathbb{E}}\bigotimes_{l=1}^k |\psi_{i_l}\rangle\langle \psi_{i_l}|\right].
\end{equation*}
The operator $\mathbb{E}_{\psi_1,\ldots,\psi_k}\bigotimes_{l=1}^k |\psi_{i_l}\rangle\langle \psi_{i_l}|$ is a tensor product over maximally mixed states on symmetric subspaces. In the following, for a subset $L\in[k]$, we denote by $\mathcal{S}_{L}$ the projector onto the symmetric subspace of tensor copies in $L$. 
If the subset is clear from context, we denote the projector onto the symmetric subspace of $a$ tensor copies by $\mathcal{S}_a$.
\begin{equation*}
    \underset{\psi_1,\ldots,\psi_k\sim\mu}{\mathbb{E}}\bigotimes_{l=1}^k |\psi_{i_l}\rangle\langle \psi_{i_l}|=\bigotimes_{q=1}^r \frac{\mathcal{S}_{\{l;i_l=q\}}}{{|\{l;i_l=q\}|+d-1\choose d-1}}.
\end{equation*}
Consequently, we find
\begin{align}
\begin{split}\label{eq:calculating bob}
\rho_{\mathrm{Bob}} &= \frac{1}{r^{2k}}\sum_{i_1,\ldots,i_k,j_1,\ldots,j_k}\underset{\psi_1,\ldots,\psi_r}{\mathbb{E}} \tr\left[A_t\bigotimes_{l=1}^k|\psi_{i_l}\rangle\langle\psi_{i_l}|\right] \bigotimes_{l=1}^k|\psi_{j_l}\rangle\langle\psi_{j_l}|\\
&=\frac{1}{r^{2k}}\sum_{i_1,\ldots,i_k,j_1,\ldots,j_k}\underset{\psi_1,\ldots,\psi_r}{\mathbb{E}}\tr\left[\bigotimes_{q=1}^{r}\mathcal{S}_{\{l;i_l=q\}} A_t\bigotimes_{q=1}^{r} \mathcal{S}_{\{l;i_l=q\}}\bigotimes_{l=1}^k|\psi_{i_l}\rangle\langle\psi_{i_l}|\right] \bigotimes_{l=1}^k|\psi_{j_l}\rangle\langle\psi_{j_l}|\\
&=\frac{1}{r^{2k}}\sum_{i_1,\ldots,i_k,j_1,\ldots,j_k}\tr\left[A_t\bigotimes_{q=1}^{r} \mathcal{S}_{\{l;i_l=q\}}\right]\underset{\psi_1,\ldots,\psi_r}{\mathbb{E}}\tr\left[\tau^t_{i_1,\ldots,j_k}\bigotimes_{l=1}^k|\psi_{i_l}\rangle\langle\psi_{i_l}|\right] \bigotimes_{l=1}^k|\psi_{j_l}\rangle\langle\psi_{j_l}|,
%\label{eq: last line of bob}
\end{split}
\end{align}
where we define the state
\begin{equation*}
\tau^t_{i_1,\ldots,j_k}:=\frac{\bigotimes_{q=1}^{r} \mathcal{S}_{\{l;i_l=q\}}A_t\bigotimes_{q=1}^{r} \mathcal{S}_{\{l;i_l=q\}}}{\tr\left[A_t\bigotimes_{q=1}^{r} \mathcal{S}_{\{l;i_l=q\}}\right]}. 
\end{equation*}
We can now interpret the expression $\mathbb{E}_{\psi_{1},\ldots,\psi_r} \tr\left[\tau^t_{i_1,\ldots,j_k}\bigotimes_{l=1}^k|\psi_{i_l}\rangle\langle\psi_{i_l}|\right] \bigotimes_{l=1}^k|\psi_{j_l}\rangle\langle\psi_{j_l}|$ as a tensor product of measure-and-prepare channels applied to the state $\tau$.
We apply Lemma~\ref{lemma:bound on post measurement} and arrive at the following bound:
\begin{multline*}
    \underset{\psi_1,\ldots,\psi_r\sim\mu}{\mathbb{E}}\tr\left[\tau^t_{i_1,\ldots,j_k}\bigotimes_{l=1}^k|\psi_{i_l}\rangle\langle\psi_{i_l}|\right] \bigotimes_{l=1}^k|\psi_{j_l}\rangle\langle\psi_{j_l}|\\
    \geq \prod_{q=1}^{r}\left(e^{-|\{l;i_l=q\}||\{l;j_l=q\}|/d} \frac{1}{{|\{l;i_l=q\}|+d-1\choose d-1}}\right)\bigotimes_{q=1}^r \frac{\mathcal{S}_{\{l;j_l=q\}}}{{|\{l;j_l=q\}|+d-1\choose d-1}}.
\end{multline*}
In Eq.~\eqref{eq:calculating bob} we will typically see tuples $i_1,\ldots,i_k$, such that every single value $1\leq q\leq r$ can be expected to appear roughly $k/r$ many times.
This will allow us to lower bound the contribution of $\prod_{q=1}^{r}e^{-|\{l;i_l=q\}||\{l;j_l=q\}|/d}$.
We can bound the tails of this distribution via a Hoeffding bound. 
More precisely, the probability of drawing the number $q$ greater than $m\geq k/r$ times is bounded by
\begin{equation*}
    \mathrm{Pr}[|\{l;i_l=q\}|\geq m]\leq 2e^{-2k(1/r-m/k)^2}.
\end{equation*}
In particular, setting $m=Ck/r$ for a constant $C>1$, we find via a union bound
\begin{equation}\label{eq:binomial bound}
    \mathrm{Pr}[\max_q\{|\{l;i_l=q\},\{l;j_l=q\}|\}\geq Ck/r]\leq  \sum_{q=1}^k\mathrm{Pr}[|\{l;i_l=q\}|\geq m]\leq  4k e^{-2k(1-C)^2/r^2}.
\end{equation}
Splitting the sum in Eq.~\eqref{eq:calculating bob} into the typical case and the tail, we find
\begin{align*}
    \begin{split}
    \rho_{\mathrm{Bob}}&\geq \frac1{r^{2k}}
    \sum_{\substack{i_1,\ldots,j_k\\
    \max_q\{|\{l;i_l=q\}|,|\{l;j_l=q\}|\}\leq Ck/r}}
    \tr\left[A_t\bigotimes_{q=1}^{r} \frac{\mathcal{S}_{\{l;i_l=q\}}}{{|\{l;i_l=q\}|+d-1\choose d-1}}\right]e^{-C^2k^2/rd}\bigotimes_{q=1}^r \frac{\mathcal{S}_{\{l;j_l=q\}}}{{|\{l;j_l=q\}|+d-1\choose d-1}}\\
   &= \frac1{r^{2k}}\sum_{i_1,\ldots,j_k=1}^r
    \tr\left[A_t\bigotimes_{q=1}^{r} \frac{\mathcal{S}_{\{l;i_l=q\}}}{{|\{l;i_l=q\}|+d-1\choose d-1}}\right]e^{-C^2k^2/rd}\bigotimes_{q=1}^r \frac{\mathcal{S}_{\{l;j_l=q\}}}{{|\{l;j_l=q\}|+d-1\choose d-1}}-L,
    \end{split}
\end{align*}
where we defined
\begin{equation*}
    L:=\frac{1}{r^{2k}}\sum_{\substack{i_1,\ldots,j_k\\
    \max_q\{|\{l;i_l=q\}|,|\{l;j_l=q\}|\}> Ck/r}}\tr\left[A_t\bigotimes_{q=1}^{r} \frac{\mathcal{S}_{\{l;i_l=q\}}}{{|\{l;i_l=q\}|+d-1\choose d-1}}\right]e^{-C^2k^2/rd}\bigotimes_{q=1}^r \frac{\mathcal{S}_{\{l;j_l=q\}}}{{|\{l;j_l=q\}|+d-1\choose d-1}}.
\end{equation*}
We can now apply this characterization to case 1 of the mixed DIPE (Problem~\ref{problem:DIPEconvex}):
\begin{align*}
\begin{split}
    &\underset{\rho}{\mathbb{E}}\tr[M_{A,B}\rho^{\otimes k}\otimes \rho^{\otimes k}]=\sum_t \tr\left[\tr[A_t\rho^{\otimes k}]\rho^{\otimes k} B_t\right]\\
    &\geq \sum_t\frac{e^{-C^2k^2/rd}}{r^{2k}}\sum_{i_1,\ldots,j_k=1}^r\tr\left[A_t\bigotimes_{q=1}^{r} \frac{\mathcal{S}_{\{l;i_l=q\}}}{{|\{l;i_l=q\}|+d-1\choose d-1}}\right]\tr\left[\bigotimes_{q=1}^r \frac{\mathcal{S}_{\{l;j_l=q\}}}{{|\{l;j_l=q\}|+d-1\choose d-1}}B_t\right]-\sum_t\tr[LB_t]\\
    &=e^{-C^2k^2/rd}\mathbb{E}_U\tr[M_{AB}\rho^{\otimes k}\otimes \sigma^{\otimes k}]-\sum_t\tr(LB_t)\geq e^{-C^2k^2/rd}\frac23-\sum_t\tr(LB_t)
    \end{split}
\end{align*}

The second term can be bounded by
\begin{align*}
    \sum_t\tr[B_t L_t]&\leq \frac{1}{r^{2k}}\sum_{\substack{i_1,\ldots,j_k\\
    \max_q\{|\{l,i_l=q\}|,|\{l,j_l=q\}|\}> Ck/r}}\tr\left[M_{AB}\bigotimes_{q=1}^{r} \frac{\mathcal{S}_{\{l;i_l=q\}}}{{|\{l;i_l=q\}|+d-1\choose d-1}}\otimes \bigotimes_{q=1}^r \frac{\mathcal{S}_{\{l;j_l=q\}}}{{|\{l;j_l=q\}|+d-1\choose d-1}}\right]\\
    &\leq \frac{1}{r^{2k}}\sum_{\substack{i_1,\ldots,j_k\\
    \max_q\{|\{l,i_l=q\}|,|\{l,j_l=q\}|\}> Ck/r}}1\\
    &= \mathrm{Pr}[\max_q\{|\{l;i_l=q\}|,|\{l;j_l=q\}|\}> Ck/r]\\
    &\leq 4k e^{-\Omega(kC^2/r^2)},
\end{align*}
where we used Hölder's inequality and Eq.~\eqref{eq:binomial bound}.
% \zz{$k\leq o(\sqrt{dr})=o(\sqrt{d/\mathbb{E}\tr[\rho^2]})$ and we only need to bound the probability above, right? I didn't follow the following part.}
% \jh{We need to lower bound the above probability but for this we need to argue that $\sum_t \tr(B_t L_t)$ is small. This is only true for $k\geq \sqrt{r}$ but I tried to argue below that we are done anyways if this is not the case.}
% \zz{I see, but what does Eq. 14 calculate?}
% \jh{The idea was that we don't need to restrict to the case $\max_q|\{i_l=q\}|\leq Ck/r$ to show that there is a contradiction in the regime $k\leq \sqrt{r}\leq  \sqrt{d}$ as Eq. 14 holds in any case because of the norm inequality $||.||_2\leq ||.||_1$}.
% \zz{Got it. Thanks!}
We choose $C$ large enough so that $ke^{-\Omega(kC^2/r^2)}\leq \frac1{20}$ for all $k\geq c\sqrt{r}$ for a small constant $c>0$.
Assuming that $k<c\sqrt{r}\leq c\sqrt{d}$, we can show that the convex DIPE is strictly harder than the pure version.
Indeed, 
\begin{equation*}
    \prod_{q=1}^{r}e^{-|\{l;i_l=q\}||\{l;j_l=q\}|/d}= e^{-\sum_q |\{l;i_l=q\}||\{l;j_l=q\}|/d}\geq e^{-k^2/d}.
\end{equation*}
Else, we simply observe that $k\leq o(\sqrt{dr})=o(\sqrt{d/\mathbb{E}\tr[\rho^2]})$ contradicts the assumption $\mathbb{E}_{U} \tr[M_{A,B}\rho^{\otimes k}\otimes \rho^{\otimes k}]\leq \frac13$, which completes the proof.
\end{proof}

%%%%%%%%%%%%%%%%%%%%%%%%%%%%%%%%%%%%%%%%%%%%%%%%%%%%%%%%%%%%%%
\section*{Acknowledgement}
We thank Anurag Anshu and Sitan Chen for illuminating suggestions and comments throughout this research. We thank Aayush Karan, Yunchao Liu, and Yihui Quek for the helpful discussions on inner product estimation and representation theory. We thank Rui Sun for helpful discussions on purity estimation. We thank Zhenhuan Liu for helpful discussions on random induced states.
JH acknowledges funding from the Harvard Quantum Initiative postdoctoral fellowship. QY was supported
in part by the National Natural Science Foundation of China Grant No.T2225008, Tsinghua University Dushi Program, and Shanghai Qi Zhi Institute.

%%%%%%%%%%%%%%%%s%%%%%%%%%%%%%%%%%%%%%%%%%%%%%%%%%%%%%%%%%%%%%
\bibliographystyle{MyRefFont}
\bibliography{inner_product_purity}
\clearpage
\appendix
%%%%%%%%%%%%%%%%%%%%%%%%%%%%%%%%%%%%%%%%%%%%%%%%%%%%%%%%%%%%%
\end{document}